\documentclass[12pts,sloppy]{article}

\usepackage{scicite}

\usepackage{times}
\usepackage{graphicx}
\usepackage{hyperref}

\newenvironment{sciabstract}{%
\begin{quote} \bf}
{\end{quote}}

\usepackage{xcolor}

\newcommand{\reffig}[1]{Fig.~\ref{#1}}

\newcommand{\vdWSph}{vdW-Spheres}

\title{Automated Structure Discovery in Atomic Force Microscopy}

\author
{Benjamin Alldritt,$^{1\ast}$ Prokop Hapala,$^{1\ast}$ Niko Oinonen,$^{1\ast}$ \\
	Fedor Urtev,$^{1,2\ast}$ Ondrej Krejci,$^{1}$ Filippo Federici Canova,$^{1,3}$ \\
	Juho Kannala,$^{2}$ Fabian Schulz,$^{1\dagger}$ Peter Liljeroth,$^{1\ddagger}$ Adam S. Foster$^{1,4,5\ddagger}$\\
\\
\normalsize{$^{1}$Department of Applied Physics, Aalto University, 00076 Aalto, Espoo, Finland}\\
\normalsize{$^{2}$Department of Computer Science, Aalto University, 00076 Aalto, Espoo, Finland}\\
\normalsize{$^{3}$Nanolayers Research Computing Ltd, London, UK}\\
\normalsize{$^{4}$Graduate School Materials Science in Mainz, Staudinger Weg 9, 55128, Germany} \\
\normalsize{$^{5}$WPI Nano Life Science Institute (WPI-NanoLSI), Kanazawa University,}\\
\normalsize{Kakuma-machi, Kanazawa 920-1192, Japan}\\
\\
\normalsize{$^\ast$These authors contributed equally.}\\
\normalsize{$^\dagger$Present address: IBM Research–-Zurich, }\\
\normalsize{S\"aumerstrasse 4, 8803 R\"uschlikon, Switzerland}\\
\normalsize{$^\ddagger$To whom correspondence should be addressed; }\\
\normalsize{E-mail: peter.liljeroth@aalto.fi; adam.foster@aalto.fi.}
}

\date{}

\begin{document} 


\baselineskip24pt


\maketitle


\begin{sciabstract}
Atomic force microscopy (AFM) with molecule-functionalized tips has emerged as the primary experimental technique for probing the atomic structure of organic molecules on surfaces. Most experiments have been limited to nearly planar aromatic molecules, due to difficulties with interpretation of highly distorted AFM images originating from non-planar molecules. Here we develop a deep learning infrastructure that matches a set of AFM images with a unique descriptor characterizing the molecular configuration, allowing us to predict the molecular structure directly. We apply this methodology to resolve several distinct adsorption configurations of 1S-camphor on Cu(111) based on low-temperature AFM measurements. This approach will open the door to apply high-resolution AFM to a large variety of systems for which routine atomic and chemical structural resolution on the level of individual objects/molecules would be a major breakthrough.
\end{sciabstract}

% In setting up this template for *Science* papers, we've used both
% the \section* command and the \paragraph* command for topical
% divisions.  Which you use will of course depend on the type of paper
% you're writing.  Review Articles tend to have displayed headings, for
% which \section* is more appropriate; Research Articles, when they have
% formal topical divisions at all, tend to signal them with bold text
% that runs into the paragraph, for which \paragraph* is the right
% choice.  Either way, use the asterisk (*) modifier, as shown, to

\section*{Introduction}

Scanning Probe Microscopy (SPM) has been the engine of characterization in nanoscale systems \cite{Loos:2005ew}. Atomic Force Microscopy (AFM) \cite{Binnig:1986vr} in particular has developed into a leading technique for high-resolution studies without material restrictions \cite{Giessibl:2003ug,Morita:2015jp,Pavlicek:2017ji}. It is increasingly being used for detailed characterization in a wide variety of physical, biological and chemical processes \cite{Mueller:2008en,Dufrene:2017gm}. Pioneering experimental studies are now providing atomic scale insight into, for example, friction, catalytic reactions, electron transport and optical response. In general for AFM, the tip itself has often been the barrier to translating atomic resolution into physical understanding, with many images and processes ultimately being identified as a convolution with the tip structure \cite{Hofer:2003gq,Barth:2010fw}. While many partially successful efforts in tip functionalization were attempted in the last decade, the use of a CO molecule attached to a metal tip in low-temperature ultra-high vacuum AFM (CO-AFM) measurements \cite{Pavlicek:2017ji,Gross:2009td} has offered a path to reliable, outstanding resolution. The use of a relatively inert tip, with respect to the molecule-substrate interaction \cite{sweetman_intramolecular_2014}, means that it can approach very close to the object of interest without excessive attractive forces resulting in unintentional lateral manipulation of the target molecule. This allows the interaction to be dominated by extremely short-ranged Pauli repulsion between atoms in the sample and at the tip apex, providing the very high resolution essential to the technique. In particular, CO-AFM now offers an unprecedented window into molecular structure on surfaces -- aside from the detailed resolution of the results of molecular assembly \cite{gross_recent_2011,Hamalainen:2014ij}, it is possible to study bond order \cite{Gross:2012jm}, charge distributions \cite{Gross:2009charge,Mohn:2012gh} and the individual steps of on-surface chemical reactions \cite{deOteyza:2013hh,Kawai:2016jl,Kawai:2017bp,Schulz:2017if}.

As yet, most CO-AFM studies have been focused on planar molecular systems, where the experimental image requires almost no interpretation \cite{Gross:2009td,Pavlicek:2017ji,Jelinek:2017gx}. Even where understanding is not immediately obvious, such as due to controversies over the nature of observed bonds \cite{Jarvis:2015kw}, efficient models have been developed \cite{Boneschanscher:2014corr,Hamalainen:2014ij,Hapala:2014kr,vanderLit:2016jg,Ellner:2019dw} that explain the contrast mechanism in terms of the tip-surface interaction and CO lateral flexibility. However, the further the systems studied are from two-dimensional molecules containing only hydrogen and carbon, the more complex and time consuming (if not impossible) the interpretation process becomes \cite{Kawai:2016jl,Gross:2010struct,Albrecht:2015,Albrecht:2016,Schuler:2015Asphalt}. While recent measurements using rigid O-terminated copper tips makes interpreting images of flat systems even easier \cite{Monig:2015cv,Monig:2018gr}, the rigidity also means even less atoms can be characterized when moving to 3D systems - the flexibility of CO allows it to sample molecular "edges" in more detail. In recent years, CO-AFM has moved towards measuring truly unknown structures \cite{Schuler:2015Asphalt,Fatayer2018_GRL,Fatayer:2018by,Schulz2019_PROCI}, where it has overcome many of the limitations of techniques such as nuclear magnetic resonance and mass spectrometry. It is clear that this trend is going to continue, and potentially even accelerate, in particular for innovative studies, e.g. in life sciences or biochemistry \cite{Mueller:2008en,Dufrene:2017gm}, demonstrated manifestly in the first CO-AFM images of DNA \cite{Pawlak:2019eu}. Reliable interpretation of such data becomes a vast exploration through all possible molecules, configurations and imaging parameters to find agreement. This is impractical in anything beyond very simple systems, severely limiting the ultimate power of the technique.

In this work, we couple a systematic software approach with detailed experimental CO-AFM imaging to understand and predict AFM images for molecules of any size, configuration or orientation without prior knowledge of the system being studied. We use the latest modelling approaches to efficiently synthesize 3D AFM data \cite{Albers:2009hz} from 134 000 isolated molecules. These were scanned from representative directions to establish physical descriptors that characterize a series of slices through the data in a given direction. For a given series of experimental images, we then apply a deep learning infrastructure \cite{LeCun:2015dt,Kalinin:2015cx,Kalinin:2016dk,Ziatdinov:2017de} to find a descriptor match, and predict the molecular structure directly. The method is validated by comparison to a systematic CO-AFM experimental study of orientations of camphor molecules on a copper surface. This Automated Structure Discovery Atomic Force Microscopy (ASD-AFM) approach will open the door to apply high-resolution AFM to a huge variety of systems for which routine atomic and chemical structural resolution on the level of individual objects/molecules would be a major breakthrough. 

\section*{Results}

The measured signal in CO-AFM is the shift of the resonance frequency of the cantilever ($\Delta f$), which is due to the sum of all conceivable tip-sample interactions. In CO-AFM, the $\Delta f$ signal is, to a large extent, determined by the interaction of oxygen in the CO molecule and the closest atoms of the sample directly under the tip. Nevertheless, due to the lateral flexibility of the CO, the image contrast is not related to the atomic positions in a trivial fashion. We will describe a methodology that aims to invert this imaging process and yield the atomic coordinates directly from a set of measured (or simulated) $\Delta f$ data. Briefly, this involves developing an image descriptor, i.e. a 2D representation of molecular structure, that encodes the positions of the atoms in the object molecule -- this can be calculated directly if the positions are known. We train a neural network to reproduce this image descriptor directly from the $\Delta f$ data using simulated AFM images and then verify this approach using simulated images from molecules not included in the training data. Finally, we will employ experimental AFM images as a final test of the proposed methodology.

\begin{figure*}[!b]
	\centering
	\includegraphics[width=120mm]{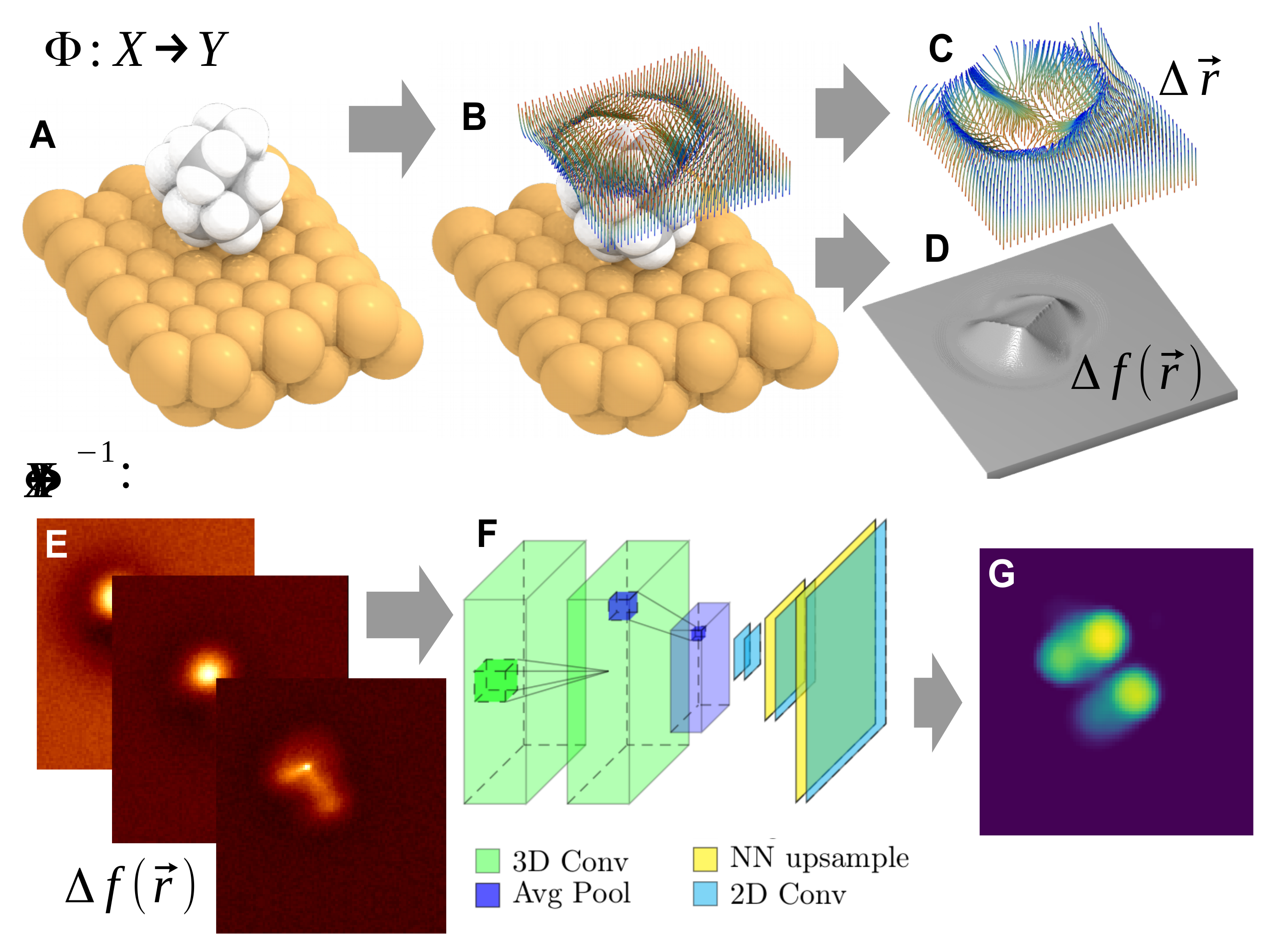}
	\caption{{\bf Schematic illustration of the CO-tip AFM imaging process and the proposed solution for the inverse imaging problem.} (\textbf{A-D}) The imaging process $ \Phi : X \rightarrow Y $ of molecular geometry $X$ (panel (A)) originates predominantly from probe particle (PP) displacement due to interactions with sample atoms (panel (B)). The resulting PP displacement $\Delta \vec{r}$ is plotted in panel (C). The fibers show deflection of the PP as it approaches toward the surface, with the red-blue gradient representing the tip-sample distance (red=far,blue=close). (\textbf{D}) The resulting AFM frequency shift ($\Delta f(\vec{r})$) images $Y$  obtained by integrating the forces felt by the relaxed PP over its path. (\textbf{E-G}) The inverse imaging process (i.e. reconstruction of geometry) $ \Phi^{-1} : Y \rightarrow X $ approximated by a convolutional neural network (F) transforming a 3D stack of AFM images $Y$ (E) to a description of the molecular geometry $X$ (represented by e.g. van der Waals spheres (G)).}
	\label{figSchematics}
\end{figure*}
\subsection*{Inverse imaging problem}
Reconstruction of molecular structures from AFM images can be seen as the search for an inverse function ($\Phi^{-1}$) to the imaging process $ \Phi : (\vec R,Z) \rightarrow {\Delta f(\vec r) } $, where $\vec R,Z$ are positions and atomic number of nuclei, and $\Delta f(\vec r)$ is the value of measured frequency shift in each point of space $\vec r$ (see \reffig{figSchematics}). Analysis and understanding of the imaging process $\Phi$ are therefore crucial for obtaining ($ \Phi^{-1}$). In particular, it is important to estimate how well conditioned the inverse operation is, and to identify which information is preserved or where information is lost.

The imaging process can be decomposed into the following sequence of operations: 

\begin{enumerate}
	\item Atoms of the sample generate various force fields in the space around them (e.g. electrostatic, van der Waals, Pauli repulsion). Many methods ranging from empirical potentials (e.g.\cite{OPLS}) to \emph{ab initio} calculations (e.g.\cite{Moll:2010}) were applied in the past to approximate those force fields.
	\item The tip apex (e.g. CO molecule) relaxes under the influence of those force fields as it approaches toward the sample (see \reffig{figSchematics}B). This means that the force fields are sampled in distorted (relaxed) coordinates (\reffig{figSchematics}C). These distortions are crucial for understanding features in AFM images. The process can be simulated by a simple mechanical model (e.g. probe particle (PP) model \cite{Hapala:2014kr,Jelinek:2017gx}).
	\item Forces felt by the relaxed probe particle are integrated over its path (\reffig{figSchematics}C) and this causes changes in the measured oscillation frequency (\reffig{figSchematics}D). The change of frequency $\Delta f$ can be therefore calculated using a simple formula \cite{Giessibl:2001formula}.
\end{enumerate}

Furthermore, from previous simulations of the AFM imaging process \cite{Hapala:2014kr,vanderLit:2016jg,Jelinek:2017gx,Peng:2018WeakH2O,Hamalainen:2014ij}, it is clear that images are extremely sensitive to even minor variations of height (z-coordinate) of the topmost atoms, and conversely very insensitive to atoms $>$0.5 \AA{} below this. Also, the chemical identity of the atom cannot be easily determined from observed contrast as it depends on the $z$-coordinate, the chemical neighborhood and orbital structure (e.g. nitrogen can appear both as a depression and a protrusion in carbonaceous aromatic systems). Instead, the characteristic topology of interatomic potentials (saddle ridges between nearby atoms, vertexes between those ridges, contrast inversion) can be clearly determined from AFM data as a fingerprint of typical chemical groups or bonding configurations. The electrostatic force has a rather small contribution to vertical force in contact, but often considerably distorts the image laterally \cite{vanderLit:2016jg,Hapala:2016NatComm}. 

Overall, the imaging process ($\Phi$; \reffig{figSchematics}A-D) is a complex and highly non-linear function, and its inversion ($\Phi^{-1}$) cannot be easily expressed by any analytic equation or practical numerical algorithm. Hence, we employ a neural network (NN) (\reffig{figSchematics}F) as an efficient universal fitting scheme to learn an approximation to $\Phi^{-1}$ from example atomic structures and corresponding 3D AFM data stacks (a stack is a set of constant height images at different vertical positions; \reffig{figSchematics}e). The image-like structure of input AFM data calls for the use of a deep convolutional neural network (CNN)\cite{LeCun:2015dt}, optimized for machine learning (ML) of regular 3D grids. 

\subsection*{Generation of training data}

The main problem in training deep convolutional networks is to provide sufficient labeled training data (from thousands to millions of input-output pairs). High-resolution AFM experiments are time intensive, requiring several hours to acquire a single 3D data stack, which would render direct training on experimental data impractical. In addition, experimental data are \emph{a priori} unlabeled (i.e. we do not know the correct interpretation) and interpretation of 3D features in AFM data is currently a difficult task, even for human experts. Hence, human labeling cannot provide us with reliable labels.

Therefore, the only feasible option is to train a model on simulated data, where correct interpretation (labels) are known \emph{a priori}. For our reference simulations, the geometries of sample molecules were taken from a well-known database of 134 000 isolated small organics~\cite{Ramakrishnan:2014ij}, structurally optimised with Density Functional Theory (DFT).

Our methodology employs a new, highly efficient graphical processing unit (GPU) implementation of the PP model \cite{Hapala:2014kr,Hapala:2014IETS}, which allows the generation of $\sim$50 input-output pairs (i.e. 3D AFM data-stacks and 2D image representation of structure) per second. This implementation is performance optimized, allowing for rapid experimentation with new settings and CNN architectures, while simultaneously generating data on-the-fly. This eliminates issues related to the storage of terabytes of training data otherwise needed. For each molecule, we first calculate the force field sampled on a regular 3D grid (this step takes $\sim$0.1s on a desktop computer), and then this force field can be rapidly interpolated to generate simulated constant-height $\Delta f$ images from 10-20 orientations of a given molecule (dependent on molecule symmetries) each of which takes $\sim$0.02s. These orientations are initially uniformly distributed over a sphere, but we then weight the final selection to orientations which expose more atoms to the tip. This avoids images where just a single atom is visible and increases the information available per stack in the training process. Here, and in general, the $z$-coordinate is defined as the distance from the carbon in the CO-tip apex to the atom closest to the tip in a particular molecular orientation. Each scan starts at $z=8.0$ \AA\ and continues 3.0 \AA\ toward the molecule in steps of 0.1 \AA. These 30 slices of vertical force are transformed into 20 slices of frequency shift (2.0 \AA\ of valid data) using the Giessibl formula \cite{Giessibl:2001formula}, forming a \emph{stack} from simulated data. Optimization of this choice of $z$-window is possible for a given experiment, but this selection provided the best performance for the results presented here.

\subsection*{Image descriptors}

In general when trying to predict molecular geometries from AFM images, while it may seem most obvious to directly convert an image stack to a set of $xyz$ coordinates, this is not an efficient descriptor in a CNN model (see expanded discussion in Supplementary Material, SM). Hence, we opt to represent the output geometry in an image-like form that is directly related to the atomic coordinates. The selection of this 2D image descriptor is critical to an efficient model and must be chosen such that it can be realistically and reliably determined from AFM data. The descriptor can be considered as the \emph{language} with which we wish to analyze the problem and the choice of language is enforced by the reference database - during the generation of the simulated image database we also calculate 2D image descriptors for all molecules and orientations.

Then we ask the CNN to \emph{translate} the data stacks into this language. It achieves this by extracting \emph{features} in a given $\Delta{f}$ slice as a function of their character and position. It does this simultaneously for all given $\Delta{f}$ slices in a data stack - features which appear in multiple slices are much more likely to be identified as important. As the \emph{deep} CNN moves through its multiple layers (\reffig{figSchematics}f), it filters these features according to the chosen biases and weights (manually optimized in this work, see SM), ultimately identifying a critical feature map. The CNN then begins the second half of its job, building a 2D image descriptor from this feature map. Using the reference database for that descriptor, it makes a prediction of the best match for a given feature. 

We designed several physically meaningful representations of molecular structure on a grid, with specifics of AFM microscopy in mind (see discussion in SM). In all cases, we represent the data as a single 2D image with the same lateral resolution as the input AFM data, which simplifies the computational analysis and allows for quick validation via human users. For the rest of the discussion, we use the \emph{vdW-Spheres} representation -- an intuitive representation of molecular structure by their van der Waals radii, commonly used in chemical visualization programs. For each molecule and orientation, we calculate the \emph{vdW-Spheres} descriptor from the reference database as follows: we calculate the van der Waals radius of all atoms and then plot this in 2D using a~$z$-range starting from position of the highest atom to 1.5~\AA{} below it, i.e.~contributions below this are ignored. The relative height of atoms in this window is represented by the their brightness in the 2D image descriptor.

\begin{figure*}[!t]
	\centering
	\includegraphics[width=\columnwidth]{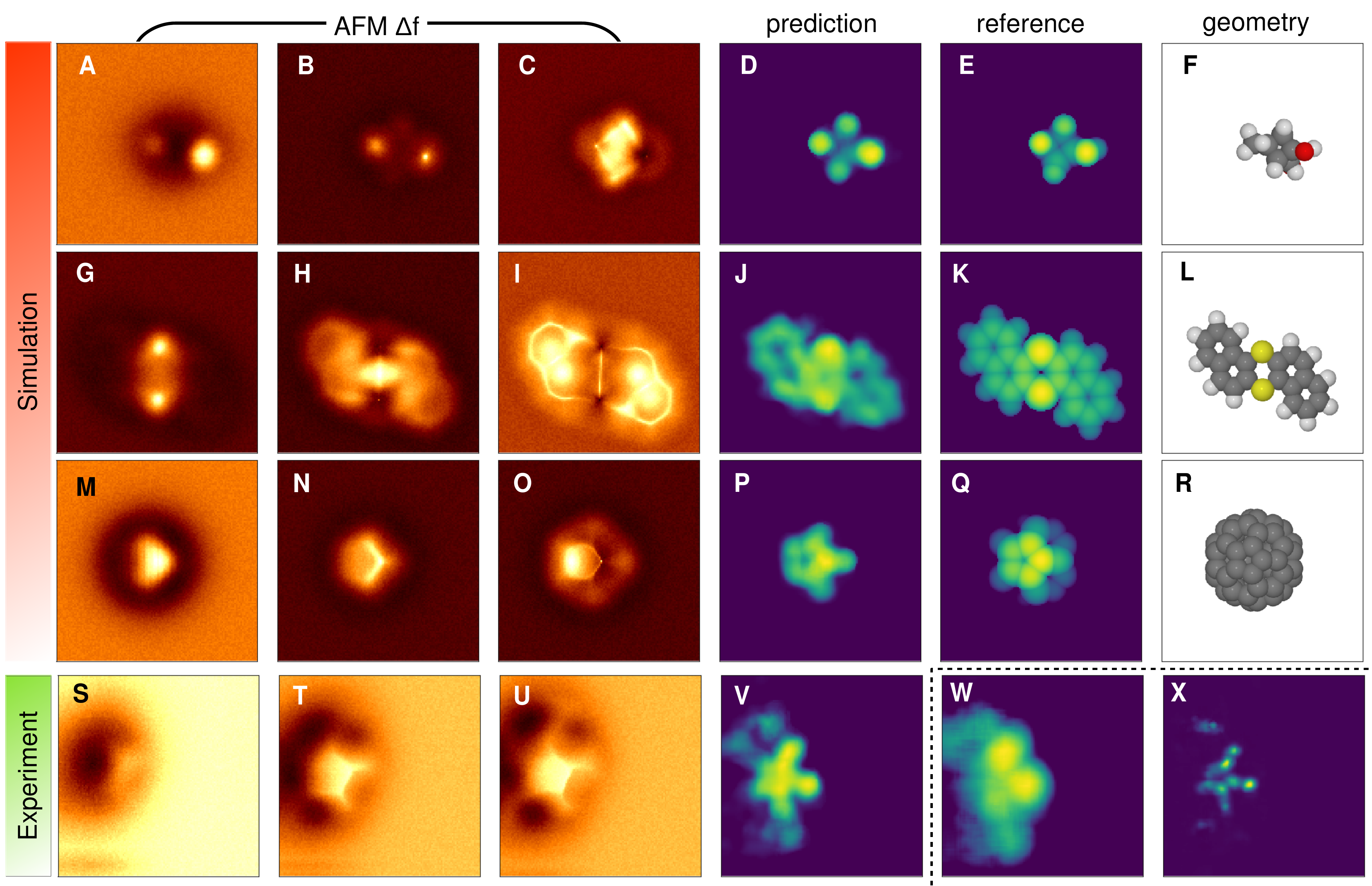}
	\caption{ { \bf Examples of CNN prediction from simulated and experimental data.} (\textbf{A-F}) A molecule from the validation set with formula C$_7$H$_{10}$O$_2$. (\textbf{G-L}) A Dibenzo[a,h]thianthrene molecule \cite{Pavlicek:2012thiarene}. (\textbf{M-U}) A fullerene C$_{60}$ (experimental data in (S-U)). (\textbf{V-X}) Comparison of image descriptors, \vdWSph{}, Height Map and Atomic Disk representation (see SM for explanation) predicted from experimental images of C$_{60}$. Columns \textbf{1-3} shows simulated AFM signal ($\Delta f$) at different heights. Column \textbf{4} shows the \vdWSph{} representation predicted by the trained CNN (naturally, the reference is not available for experiment). Column \textbf{5} shows the reference \vdWSph{} representation calculated directly from geometry. Column \textbf{6} depicts a 3D render of the molecule.
	}
	%\label{figRepresentations-s}
	\label{figFamous}
\end{figure*}

\subsection*{Geometry prediction from simulated AFM data}

In order to benchmark the methodology, we employed the trained CNN model to predict the geometry of several molecules that were not included in the training set. The \emph{internal} quality of the model can be judged by how well the predicted 2D image descriptor (derived from the simulated AFM 3D image stack) matches the reference descriptor calculated directly from the molecular geometry. In the first example (\reffig{figFamous}A-F), we picked a molecule (an isomer of C$_7$H$_{10}$O$_2$) that has a functional group and a non-planar geometry as representative of the types of molecule we wish to identify. The prediction qualitatively matches the reference, capturing all the key atoms except the hydrogen of the hydroxyl group, which is present in the analytically computed reference image representation. It is very difficult to identify the lower lying atoms from the AFM images. For the molecule shown in \reffig{figFamous}A-F, it would not be possible for a human expert to identify the hydrogen atom of the hydroxyl group. The goal of the introduced \emph{ideal} image representation, i.e. \vdWSph \ representation, is to train a CNN to extract as much as possible structural information presented in an individual AFM stack of data and store it in compressed readable format.

As another example, we consider a dibenzo[a,h]thianthrene molecule, which has been previously experimentaly studied \cite{Pavlicek:2012thiarene} (\reffig{figFamous}G-L). The CNN is again able to predict most of molecular features in the \vdWSph \ representation, in particular, identifying the two dominant sulphur atoms. The remaining atoms of the aromatic system are also predicted, but they are not as well separated as in the reference. CNN-predicted properties are typically blurred and this is somewhat dependent on the choice of 2D image descriptor (see Fig. S3g).

The last example is a fullerene C$_{60}$ molecule oriented with a pentagon upwards. We performed a prediction of the \vdWSph \ representation based both on simulation (\reffig{figFamous}M-R) and newly measured experimental data (\reffig{figFamous}S-V). The pentagons are oriented slightly in an asymmetric manner with 3 carbon atoms up. The main features, i.e. 8 top-most atoms, are reproduced rather well in the CNN prediction, while the remaining atoms remain invisible. This is true for both simulated and experimental images. In the experimental image, however, are visible artifacts originating from dark attractive areas of C$_{60}$, which are not visible in the simulated image. This is a clear indication that the simulation does not reproduce this particular experiment sufficiently well. Despite this fact, the CNN prediction is robust enough to consistently render the top-most atoms. More examples from our training set can be found in Fig.~S4.
\begin{figure*}[!bh]
	\centering
	\includegraphics[width=\columnwidth]{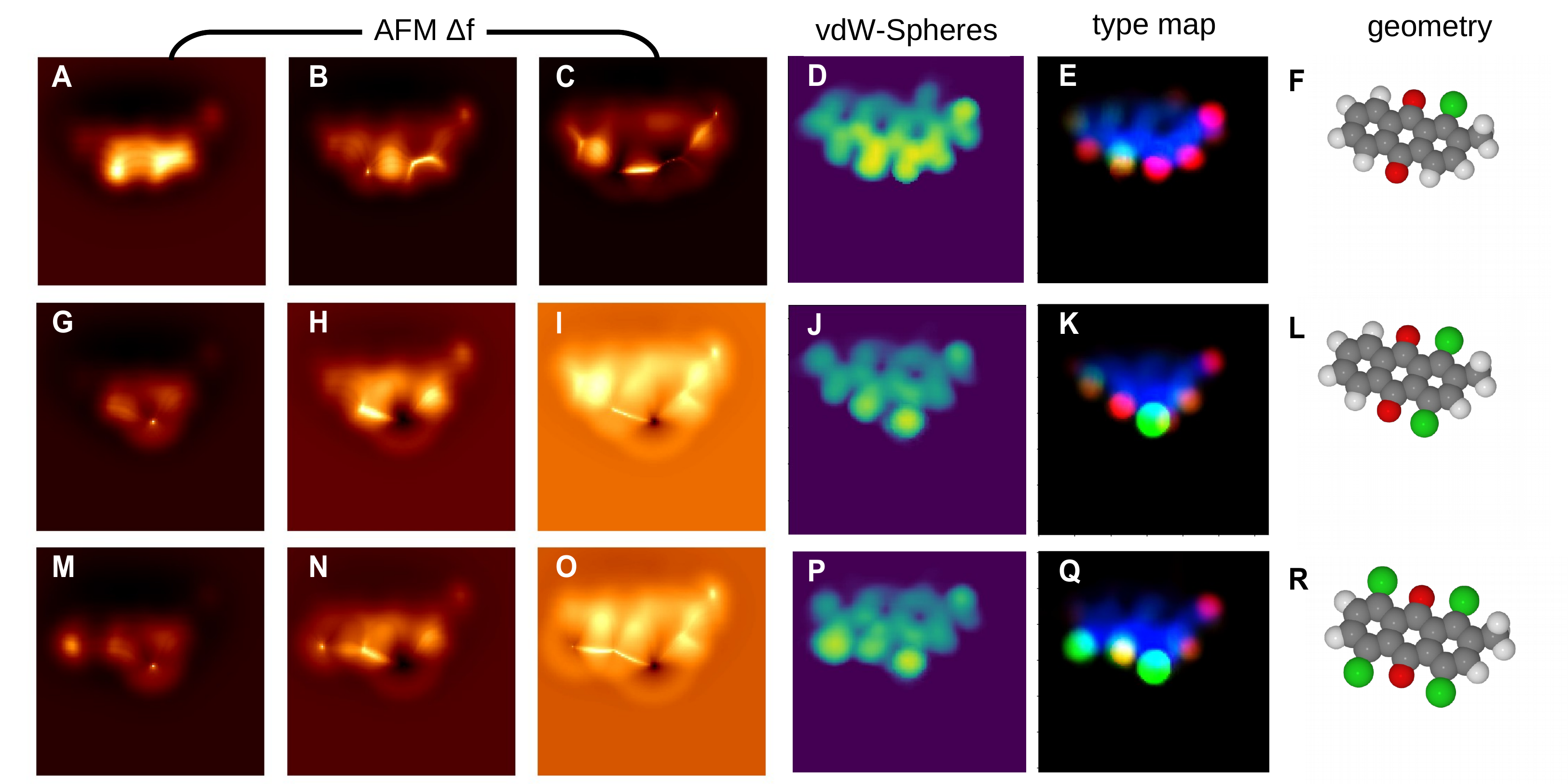}
	\caption{ {\bf Discrimination of functional groups.} Here we compare three hypothetical anthraquinone derivatives with differing numbers of chlorine atoms. The first three columns show simulated AFM images at far, middle and close tip-sample distances. The fourth column shows the associated NN prediction for the \vdWSph \ representation. 
		The 5th column shows atom type prediction from another NN that discriminates 3 different types of atoms: hydrogens (red), non-hydrogen peripheral (green) and carbon backbone (blue). The final column shows the molecular geometry. Note that the molecule is tilted so that the bottom edge is higher than the upper edge. 
	}
	\label{figComp}
\end{figure*}

To illustrate how our method can aid in discrimination of unknown molecules and separate chemical information and physical topography, we compare 3 different derivatives of antraquinone with a different number of chlorine atoms in \reffig{figComp}. In this illustrative example, the molecules are tilted so that the bottom edge is higher than the upper edge, making this a 3D problem with a peculiar image contrast over the edge that can hardly be deciphered by an expert. Although each molecule provides clearly distinct AFM images, it is rather difficult to rationalize the differences in terms of atomic structure. In fact any similarity between molecules in the the 1st and 2nd row is hardly visible from the AFM pictures. In contrast, the predicted vdW-Spheres map clearly shows a change in atomic radius in one or two atomic sites while the rest of the molecular structure is preserved. While disentangling the atomic type from its $z$-position is difficult based on the \vdWSph~image description, the different atomic types should result in a different decay of the $\Delta f$ contrast as a function of the tip-sample distance. Hence, it should be possible to differentiate atomic species. In particular, a modified CNN (shown in \reffig{figComp} as column {\it type map}) learned to discriminate small peripheral atoms (hydrogen, red) from larger peripheral atoms (chlorine, oxygen, green), leaving aside rather indiscriminate carbon backbone (blue). The network clearly identified substitution of a hydrogen atom by chlorine. While showing the potential of the technique in terms of recognition, the prediction is not yet fully reliable, as can be see from  misidentified oxygen as small (red) in the second row.

\subsection*{Geometry prediction from experimental AFM data}

\begin{figure*}[!th]
	\centering
	\includegraphics[width=\columnwidth]{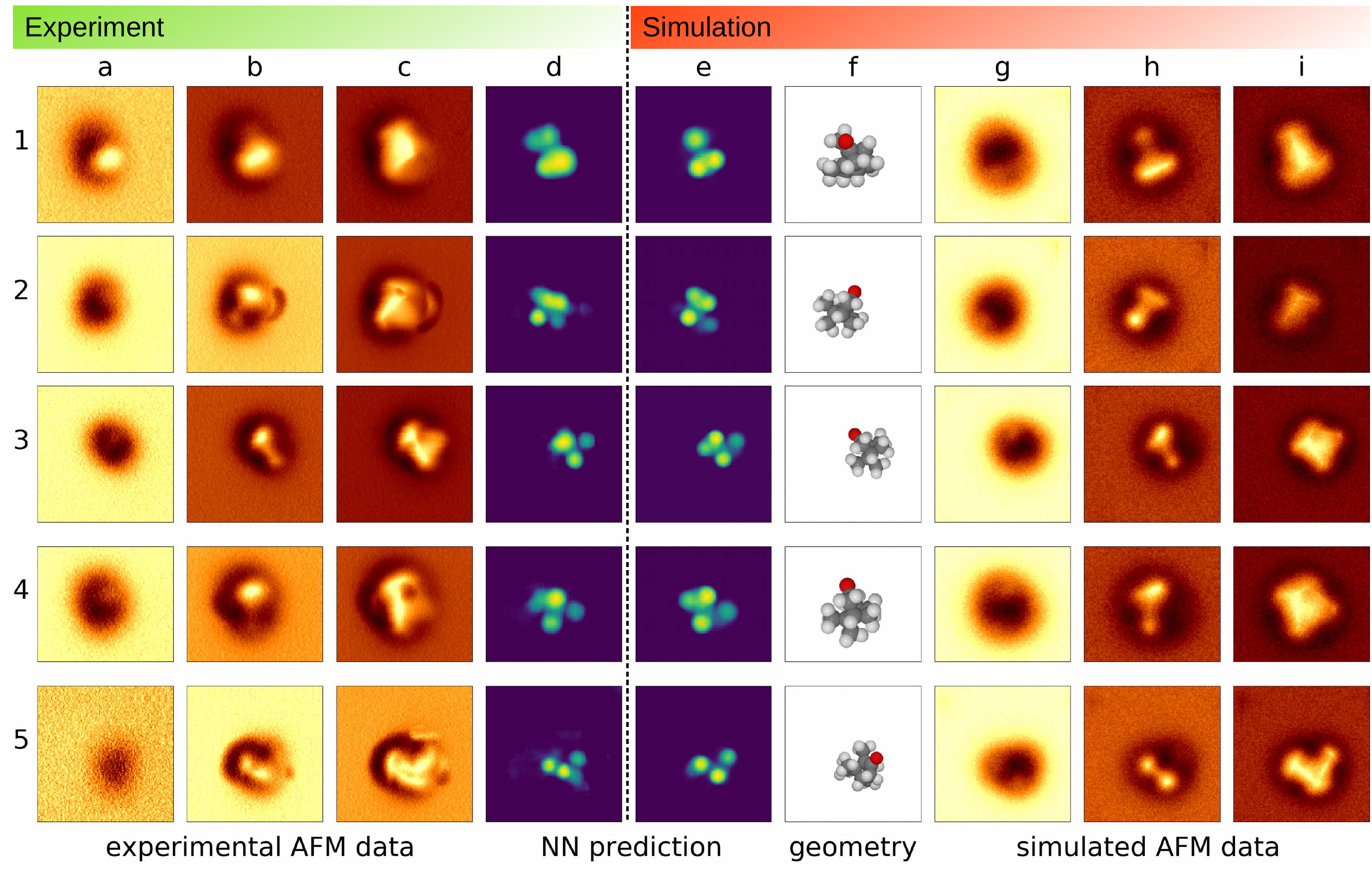}
	\caption{ { \bf Identification of the 1S-camphor adsorption configurations on Cu(111) with ASD-AFM.} \textbf{1-5} refer to distinct molecular configurations with experiments in columns (\textbf{A-D}) and simulations in columns (\textbf{E-I}). Selected experimental AFM images (out of 10 slices used for input): at (A) -- far, (B) -- middle, (C) -- close tip-sample distances and NN prediction (D) for the \vdWSph \   representation. The \vdWSph \ representation shown in (E) corresponds to the full molecular configuration (F) resulting from the best match to experiment. The corresponding simulated AFM images are given in panels (G-I) (far--middle--close).}
	\label{figExp}
\end{figure*}

The true validation of our ML approach is to make predictions directly from experimental AFM images. Ultimately, this would be done from images of an unknown system, but as a benchmark for our first iteration of the method, we apply it to find molecular configurations of a known molecule. Here we selected 1S-camphor as the target molecule due to its 3D geometry and potential for adopting multiple distinct adsorption geometries on a Cu(111) surface. Combined STM and AFM imaging allowed us to distinguish 8 characteristic adsorption geometries with reproducible data in each case. Further analysis reduced this to a set of 5 distinct configurations clean enough for good comparison and we acquired a set of constant-height $\Delta f$ images in each case (see SM for details). Even highly trained experts were not able to decipher the molecular structure from these images, and they provided an excellent challenge and example for the CNN model. The 3D experimental image stack (\reffig{figExp}A-C) is fed into the CNN model and a 2D image descriptor (\vdWSph) is predicted based on this data (\reffig{figExp}D). This \emph{experimental} descriptor is then compared via cross-correlation to a set of descriptors calculated directly from atomic coordinates taken from a set of uniformly distributed molecular rotations (\reffig{figExp}E). The best fit gives us a prediction of the molecular configuration corresponding to the original descriptor from experimental data (\reffig{figExp}F). Qualitatively, the match between experimental and simulated descriptors is very good, reproducing the performance seen with purely simulated data (\reffig{figFamous}). In order to explore the plausibility of the predicted geometries, we now reverse the inverse imaging process and consider the predicted simulated images for the best fit descriptor (\reffig{figExp}G-I). In all cases the simulated images qualitatively capture the main features seen in the experimental images. In cases 1-4, agreement is generally good at all heights, but the simulated image tends to be somewhat sharper than the experiments at close approach. For case 5, the core of the simulated image is representative of experiments, but some of the extended features are clearly absent. Furthermore, note that experimental image 5a in \reffig{figExp} shows no atomic features (the interactions are purely attractive), whereas the simulated image 5G clearly does (showing the onset of repulsive short-range interactions). This is because the CNN was consciously trained only on data containing atomic-like features, as those are critical for identification, and not the kind of large tip-sample distance used in 5A. 

\section*{Discussion}
The aim of this work was to establish a reliable and rapid method for solving a problem that expert humans cannot - the interpretation of high-resolution AFM images of complex 3D molecules. We have demonstrated that our ML method based on a CNN architecture can solve this problem with trivial computational effort. In its current form, the model can, \emph{e.g.}, identify adsorption configurations accurately. On a complex system, this allows us to drastically reduce the number of possible molecular solutions from a set of experimental images.

However, we believe this is only the first step in a developing analysis field and it is clear that several further problems need to be tackled if we wish to increase prediction accuracy even further. Simple improvements include introducing a bigger variety of atoms into the training set (with a very large initial computational cost), and the creation of an integral model that can predict multiple 2D image representations simultaneously, improving model robustness for features recognition. In the medium term, while our current approach using the PP method (i.e. re-using a precalculated force-field grid for scans from multiple directions) is highly efficient, it prevents a simple implementation of more sophisticated non-spherical electrostatics (e.g. quadrupoles) that have been shown to be important for CO tip simulations in certain systems \cite{Peng:2018WeakH2O,Ellner:2016ElecCO}. While we consider this limitation of the underlying simulation model a secondary issue in the development of a reliable ML architecture, we have already begun exploring efficient solvers for more sophisticated models based on the electron density from DFT \cite{Ellner:2019DensOver}. A more pressing concern for accuracy in simulated images is the role of surface- and tip-induced molecular displacements. For the latter, this has generally been ignored in previous simulations of CO-tip AFM experiments, and fixed geometries are considered throughout. In this work, we considered how molecular tilting and functional group rotations affected the predicted images (see SM Sec.~3). It is clear that these can change the predicted simulated images, particularly at close approach and finding a systematic way to include these in the matching process could significantly improve accuracy. We also considered the possible changes of molecular configurations when adsorbed on the surface (see SM Sec.~2), but any errors seen were not in the predictions of CNN model and improvements would require advances beyond the standard methods used to obtain accurate adsorption structures - a separate research field. 

Finally, the nature of the AFM measurement itself causes a particular difficulty in the uniqueness of the molecular solutions. For certain configurations, common in small non-planar molecules, AFM data may provide information only about a very limited number of atoms and this may lead to several molecular solutions being almost equivalent in the quality of best fit to experiments (see SM Sec.~3). In systems where this is a problem, considering several experimental configurations of the same molecule, as done here, makes identification significantly easier. More generally, we are looking at including multiple channels of information for a single configuration by using an image descriptor incorporating tip-dependent electrostatic information available via other tip terminations \cite{Gross:2014PRB,vanderLit:2016jg,Hapala:2016NatComm}. This could be also be extended to incorporate simultaneous fitting to Kelvin Probe Force Microscopy data \cite{Melitz:2011ea,sadewasser_kelvin_2012,Sadewasser:2018tq,Schulz:2018ck}, further improving the uniqueness of predictions.

Despite these challenges, the approach is immediately applicable to a wide variety of complex molecular systems where conventional interpretation approaches have either failed or cannot even be attempted. As such, it promises the availability of atomic and chemical structural resolution in systems where it offers the prospect of major impact.

\section*{Materials and Methods}

\subsection*{Machine-Learning model architecture}

The architecture of our CNN  is similar to the encoder-decoder type networks that have been used in, for example, image segmentation \cite{Badrinarayanan:2017segnet}. At the input side it comprises 3 layers of 3D convolutional filters ($3\times3\times3$) interleaved by average pooling ($2\times2\times2$), which reduces the size of the input image by a factor of 8 in $x,y$ dimensions. This information bottleneck is motivated by the fact that input AFM images are mostly rather smooth and carry a limited amount of information (i.e. just position and size of a few atoms). Down-sampling also helps to facilitate long-range correlations in the image using only local and cheap $3\times3\times3$ filters. This should help to recognize larger features such as atoms and bonds spanning over tens of pixels. The data is collapsed in the z-direction from 3D to 2D by the action of the pooling layers, while gradually being expanded to several independent channels ($2\times$ channels by each layer). Therefore, the features obtained after this operation should encode varying z-dependence of the frequency shift. The signal is further processed by 3 layers of purely convolutional filters operating independently on each of 64 channels of the 2D image. In the last part of the CNN architecture, the image is expanded back to original resolution ($8\times$ in each dimension) by 3 bi-layers of 2D convolution interleaved by NN-upsample operations. The final convolution is followed by a rectified linear unit (ReLU\cite{Glorot:2011a})  activation, which basically cuts the negative part of activations from the convolution layer, leaving 'unchanged' positive values. Other convolutions are followed by LeakyReLU activations with a factor of $0.1$ on the negative side, so as not to completely block learning when values are under 0 (they are \emph{leaked} through). The model is implemented in Keras \cite{Chollet:2015keras} running a TensorFlow \cite{Abadi:2015tf} backend. Optimization of kernel sizes in the convolutional layer has not been systematically tested, but for our image recognition network, small kernel sizes with additional layers have been quite effective.

The structure was motivated by the idea that the central part - i.e. the $8\times$ down-sampled representation with 64 channels - will learn to represent AFM images in terms of abstract, physically meaningful features (e.g. slope of frequency shift curve, blobs representing atoms, characteristic sharp-line features between nearby atoms). Various physical properties, such as height maps or positions of atoms in the second up-sampling stage, can then be identified from this internal abstract representation. 

In order to make the model more robust to experimental artifacts and limitations we add $5\%$ white noise (representing electronic noise in the measurements) and random rectangular cutouts \cite{Devries2017Cutout} (representing sudden jumps in the measurements) to the simulation data. Note that this also aids in avoiding problems in relation to the ill-posed nature of the force-frequency shift conversion \cite{Dagdeviren:2018jm,Sader:2018ey}.

\subsection*{Molecular database}

The original structures of the molecules in the database were optimized with DFT at the B3LYP/6-31G level \cite{Becke:1993is}. Using the quantum chemistry software Psi4~\cite{Parrish:2017:psi4, Crawford:2007:cc}, we performed single-point coupled-cluster calculations (singles and doubles, cc-pvdz basis) for all the 134k molecules, thus obtaining charge densities and Mulliken populations necessary to operate the Probe-Particle simulator.

\subsection*{Experimental Methods}

Polished Cu(111) and Au(111) single-crystals (Mateck/Germany) were prepared by repeated Ne+ sputtering (0.75 keV, 15 mA, 20 min) and annealing (850-900 K, 5 min) cycles. Surface cleanliness and structure was verified by scanning tunneling microscopy (STM). Sample temperatures during annealing were measured with a pyrometer (SensorTherm Metis MI16). 1S-camphor (Sigma-Aldrich, purity $>$ 98.5$\%$) was introduced into the vacuum system via a leak valve and deposited onto the Cu(111) surface at a low-temperature ($T=20$ K) to increase the number of distinct adsorption configurations and to achieve individual molecules rather than clusters on the surface. Fullerene C$_{60}$ (Sigma-Aldrich, purity $>$ 99.9$\%$) was sublimed onto a Au(111) substrate held at $\sim200$ K.

The STM and CO-AFM images were taken with a Createc LT-STM/AFM with a commercial qPlus sensor with a Pt/Ir tip, operating at approximately $T = 5$\ K in UHV at a pressure of $1\times10^{-10}$ mbar. The quartz cantilever (qPlus sensor) had a resonance frequency of $f_{0}=29939$ Hz, a quality factor $Q=101099$, and was operating with an oscillation amplitude $A=50$ pm. Tip conditioning was performed by repeatedly bringing the tip into contact with the copper surface and applying bias pulses until the necessary STM resolution was achieved. The tip apex was functionalized with a CO molecule\cite{Bartels:1997} before AFM measurements. The STM images were recorded in constant-current mode, while the AFM operated in constant-height mode. Raw data was used as input for the machine learning infrastructure. In order to minimize experimental artefacts that would cause problems with interpretation, we have implemented the following measures: Checking the background $\Delta f$ before CO pickup (smaller value indicates sharper overall tip); scanning another CO to ensure the symmetry of the CO tip after tip passivation and prior to further AFM imaging; and confirming that the excitation (dissipation) signal remains flat/featureless during the AFM measurements.

% Your references go at the end of the main text, and before the
% figures.  For this document we've used BibTeX, the .bib file
% scibib.bib, and the .bst file Science.bst.  The package scicite.sty
% was included to format the reference numbers according to *Science*
% style.

%BibTeX users: After compilation, comment out the following two lines and paste in
% the generated .bbl file. 

\section*{Supplementary Material} 
Accompanies this paper at {\small {\tt http://www.scienceadvances.org/}}.\\
Section S1. Image representations of output molecular structure\\
Section S2. Matching experiment to relaxed on-surface simulated configurations\\
Section S3. Effect of small perturbations on AFM imaging and matching\\
Section S4. Neural network architecture\\
Section S5. Probe Particle simulations\\
Figure S1. Different 2D image representations of a C$_7$H$_{10}$O$_2$ molecule from the training set\\
Figure S2. Different 2D image representations of a C$_{60}$ molecule\\
Figure S3. Different 2D image representations of a Dibenzo[a,h]thianthrene molecule\\
Figure S4. Molecules from the validation data set together with the \vdWSph~representation predicted by the CNN\\
Figure S5.  Matching between simulated relaxed configurations of 1S-Camphor and experiment\\
Figure S6. Effect of tilt of molecules on simulated AFM images\\
Figure S7. Adjustment of simulated configuration by -CH$_3$ group rotations\\
Figure S8. Matching experimental configuration \textbf{2} of 1S-Camphor with closest simulated configurations\\
Figure S9. Illustration of the layers of the CNN model \\
Figure S10. The mean squared loss for height maps, \vdWSph~and atomic disks\\
References (68-81)\\

\noindent \textbf{Acknowledgements:} \\
% Acknowledgments should be gathered into a paragraph after the final numbered reference. This section should also include 
% * complete funding information, 
% * a description of each authorÕs contribution to the paper, 
% * a listing of any competing interests of any of the authors (all authors must also fill out the Conflict of Interest form), and, 
% * a section on data and materials availability, information about the location of the data if not included in the paper, including **accession numbers** to any data relating to the paper and deposited in a public database.
%
\noindent \textbf{Funding:} Computing resources from the Aalto Science-IT project and CSC, Helsinki are gratefully acknowledged. This research made use of the Aalto Nanomicroscopy Center (Aalto NMC) facilities and was supported by the European Research Council (ERC 2017 AdG no.~788185 "Artificial Designer Materials''), and the Academy of Finland (Projects no. 311012, 314862, 314882, Centres of Excellence Program project no. 284621, and Academy professor funding no.~318995 and 320555). ASF has been supported by the World Premier International Research Center Initiative (WPI), MEXT, Japan.\\
\noindent \textbf{Author Contributions} FS, PL and ASF conceived the research. PH, NO, FU, OK and FFC developed the software and ran the simulations. BA performed the experiments. All authors were involved in the results analysis and contributed to the manuscript.\\
\noindent \textbf{Competing Interests} The authors declare that they have no competing financial interests.\\
\noindent \textbf{Data and materials availability:} All data needed to evaluate the conclusions in the paper are present in the paper and/or the Supplementary Materials. Additional data available from authors upon request.


\begin{thebibliography}{10}

\bibitem{Loos:2005ew}
J.~Loos, {The Art of SPM: Scanning Probe Microscopy in Materials Science}.
\newblock {\it Adv. Mat.\/} {\bf 17}, 1821--1833 (2005).

\bibitem{Binnig:1986vr}
G.~Binnig, C.~Quate, C.~Gerber, {Atomic Force Microscope}.
\newblock {\it Phys. Rev. Lett.\/} {\bf 56}, 930--933 (1986).

\bibitem{Giessibl:2003ug}
F.~Giessibl, {Advances in atomic force microscopy}.
\newblock {\it Rev. Mod. Phys.\/} {\bf 75}, 949--983 (2003).

\bibitem{Morita:2015jp}
S.~Morita, F.~J. Giessibl, E.~Meyer, R.~Wiesendanger, eds., {\it {Noncontact
  Atomic Force Microscopy}\/}, NanoScience and Technology (Springer
  International Publishing, Cham, 2015).

\bibitem{Pavlicek:2017ji}
N.~Pavli{\v c}ek, L.~Gross, {Generation, manipulation and characterization of
  molecules by atomic force microscopy}.
\newblock {\it Nat. Rev. Chem.\/} {\bf 1}, 0005 (2017).

\bibitem{Mueller:2008en}
D.~J. Mueller, Y.~F. Dufrene, {Atomic force microscopy as a multifunctional
  molecular toolbox in nanobiotechnology}.
\newblock {\it Nat. Nanotech.\/} {\bf 3}, 261--269 (2008).

\bibitem{Dufrene:2017gm}
Y.~F. Dufrene, T.~Ando, R.~Garcia, D.~Alsteens, D.~Mart{\'\i}nez-Mart{\'\i}n,
  A.~Engel, C.~Gerber, D.~J. M{\"u}ller, {Imaging modes of atomic force
  microscopy for application in molecular and cell biology}.
\newblock {\it Nat. Nanotech.\/} {\bf 12}, 295--307 (2017).

\bibitem{Hofer:2003gq}
W.~Hofer, A.~Foster, A.~Shluger, {Theories of scanning probe microscopes at the
  atomic scale}.
\newblock {\it Rev. Mod. Phys.\/} {\bf 75}, 1287--1331 (2003).

\bibitem{Barth:2010fw}
C.~Barth, A.~S. Foster, C.~R. Henry, A.~L. Shluger, {Recent Trends in Surface
  Characterization and Chemistry with High-Resolution Scanning Force Methods}.
\newblock {\it Adv. Mat.\/} {\bf 23}, 477--501 (2011).

\bibitem{Gross:2009td}
L.~Gross, F.~Mohn, N.~Moll, P.~Liljeroth, G.~Meyer, {The Chemical Structure of
  a Molecule Resolved by Atomic Force Microscopy}.
\newblock {\it Science\/} {\bf 325}, 1110--1114 (2009).

\bibitem{sweetman_intramolecular_2014}
A.~Sweetman, S.~P. Jarvis, P.~Rahe, N.~R. Champness, L.~Kantorovich,
  P.~Moriarty, Intramolecular bonds resolved on a semiconductor surface.
\newblock {\it Physical Review B\/} {\bf 90}, 165425 (2014).

\bibitem{gross_recent_2011}
L.~Gross, {Recent advances in submolecular resolution with scanning probe
  microscopy}.
\newblock {\it Nat. Chem.\/} {\bf 3}, 273--278 (2011).

\bibitem{Hamalainen:2014ij}
S.~K. H{\"a}m{\"a}l{\"a}inen, N.~van~der Heijden, J.~van~der Lit, S.~den
  Hartog, P.~Liljeroth, I.~Swart, {Intermolecular Contrast in Atomic Force
  Microscopy Images without Intermolecular Bonds}.
\newblock {\it Phys. Rev. Lett.\/} {\bf 113}, 186102 (2014).

\bibitem{Gross:2012jm}
L.~Gross, F.~Mohn, N.~Moll, B.~Schuler, A.~Criado, E.~Guitian, D.~Pena,
  A.~Gourdon, G.~Meyer, {Bond-Order Discrimination by Atomic Force Microscopy}.
\newblock {\it Science\/} {\bf 337}, 1326--1329 (2012).

\bibitem{Gross:2009charge}
L.~Gross, F.~Mohn, P.~Liljeroth, J.~Repp, F.~J. Giessibl, G.~Meyer, Measuring
  the charge state of an adatom with noncontact atomic force microscopy.
\newblock {\it Science\/} {\bf 324}, 1428-1431 (2009).

\bibitem{Mohn:2012gh}
F.~Mohn, L.~Gross, N.~Moll, G.~Meyer, {Imaging the charge distribution within a
  single molecule}.
\newblock {\it Nat. Nanotech.\/} {\bf 7}, 227--231 (2012).

\bibitem{deOteyza:2013hh}
D.~G. de~Oteyza, P.~Gorman, Y.~C. Chen, S.~Wickenburg, A.~Riss, D.~J. Mowbray,
  G.~Etkin, Z.~Pedramrazi, H.~Z. Tsai, A.~Rubio, M.~F. Crommie, F.~R. Fischer,
  {Direct Imaging of Covalent Bond Structure in Single-Molecule Chemical
  Reactions}.
\newblock {\it Science\/} {\bf 340}, 1434--1437 (2013).

\bibitem{Kawai:2016jl}
S.~Kawai, V.~Haapasilta, B.~D. Lindner, K.~Tahara, P.~Spijker, J.~A.
  Buitendijk, R.~Pawlak, T.~Meier, Y.~Tobe, A.~S. Foster, E.~Meyer, {Thermal
  control of sequential on-surface transformation of a hydrocarbon molecule on
  a copper surface}.
\newblock {\it Nat. Commun.\/} {\bf 7}, 12711 (2016).

\bibitem{Kawai:2017bp}
S.~Kawai, K.~Takahashi, S.~Ito, R.~Pawlak, T.~Meier, P.~Spijker, F.~F. Canova,
  J.~Tracey, K.~Nozaki, A.~S. Foster, E.~Meyer, {Competing Annulene and
  Radialene Structures in a Single Anti-Aromatic Molecule Studied by
  High-Resolution Atomic Force Microscopy}.
\newblock {\it ACS Nano\/} {\bf 11}, 8122--8130 (2017).

\bibitem{Schulz:2017if}
F.~Schulz, P.~H. Jacobse, F.~F. Canova, J.~van~der Lit, D.~Z. Gao, A.~van~den
  Hoogenband, P.~Han, R.~J. M.~K. Gebbink, M.-E. Moret, P.~M. Joensuu,
  I.~Swart, P.~Liljeroth, {Precursor Geometry Determines the Growth Mechanism
  in Graphene Nanoribbons}.
\newblock {\it J. Phys. Chem. C\/} {\bf 121}, 2896--2904 (2017).

\bibitem{Jelinek:2017gx}
P.~Jel\'inek, {High resolution SPM imaging of organic molecules with
  functionalized tips}.
\newblock {\it J. Phys. Condens. Matter\/} {\bf 29}, 343002 (2017).

\bibitem{Jarvis:2015kw}
S.~P. Jarvis, M.~A. Rashid, A.~Sweetman, J.~Leaf, S.~Taylor, P.~Moriarty,
  J.~Dunn, {Intermolecular artifacts in probe microscope images of of
  ${\mathrm{C}}_{60}$ assemblies}.
\newblock {\it Phys. Rev. B\/} {\bf 92}, 241405 (2015).

\bibitem{Boneschanscher:2014corr}
M.~P. Boneschanscher, S.~K. H{\"a}m{\"a}l{\"a}inen, P.~Liljeroth, I.~Swart,
  Sample corrugation affects the apparent bond lengths in atomic force
  microscopy.
\newblock {\it ACS Nano\/} {\bf 8}, 3006-3014 (2014).

\bibitem{Hapala:2014kr}
P.~Hapala, G.~Kichin, C.~Wagner, F.~S. Tautz, R.~Temirov, P.~Jel\'inek,
  {Mechanism of high-resolution STM/AFM imaging with functionalized tips}.
\newblock {\it Phys. Rev. B\/} {\bf 90}, 085421 (2014).

\bibitem{vanderLit:2016jg}
J.~van~der Lit, F.~Di~Cicco, P.~Hapala, P.~Jel\'inek, I.~Swart, {Submolecular
  Resolution Imaging of Molecules by Atomic Force Microscopy: The Influence of
  the Electrostatic Force}.
\newblock {\it Phys. Rev. Lett.\/} {\bf 116}, 096102 (2016).

\bibitem{Ellner:2019dw}
M.~Ellner, P.~Pou, R.~P{\'{e}}rez, Molecular identification, bond order
  discrimination, and apparent intermolecular features in atomic force
  microscopy studied with a charge density based method.
\newblock {\it {ACS} Nano\/} {\bf 13}, 786--795 (2019).

\bibitem{Gross:2010struct}
L.~Gross, F.~Mohn, N.~Moll, G.~Meyer, R.~Ebel, W.~M. Abdel-Mageed, M.~Jaspars,
  Organic structure determination using atomic-resolution scanning probe
  microscopy.
\newblock {\it Nature Chem.\/} {\bf 2}, 821-825 (2010).

\bibitem{Albrecht:2015}
F.~Albrecht, N.~Pavli\v{c}ek, C.~Herranz-Lancho, M.~Ruben, J.~Repp,
  Characterization of a surface reaction by means of atomic force microscopy.
\newblock {\it J. Am. Chem. Soc.\/} {\bf 137 23}, 7424-8 (2015).

\bibitem{Albrecht:2016}
F.~Albrecht, F.~Bischoff, W.~Auw\"arter, J.~V. Barth, J.~Repp, Direct
  identification and determination of conformational response in adsorbed
  individual nonplanar molecular species using noncontact atomic force
  microscopy.
\newblock {\it Nano Lett.\/} {\bf 16}, 7703-7709 (2016).

\bibitem{Schuler:2015Asphalt}
B.~Schuler, G.~Meyer, D.~Peña, O.~C. Mullins, L.~Gross, Unraveling the
  molecular structures of asphaltenes by atomic force microscopy.
\newblock {\it J. Am. Chem. Soc.\/} {\bf 137}, 9870-9876 (2015).

\bibitem{Monig:2015cv}
H.~M\"onig, D.~R. Hermoso, O.~D. Arado, M.~Todorovi{\'c}, A.~Timmer,
  S.~Schueer, G.~Langewisch, R.~Perez, H.~Fuchs, {Submolecular Imaging by
  Noncontact Atomic Force Microscopy with an Oxygen Atom Rigidly Connected to a
  Metallic Probe}.
\newblock {\it ACS Nano\/} {\bf 10}, 1201--1209 (2016).

\bibitem{Monig:2018gr}
H.~M\"onig, {Copper-oxide tip functionalization for submolecular atomic force
  microscopy}.
\newblock {\it Chem. Commun.\/} {\bf 54}, 9874--9888 (2018).

\bibitem{Fatayer2018_GRL}
S.~Fatayer, A.~I. Coppola, F.~Schulz, B.~D. Walker, T.~A. Broek, G.~Meyer,
  E.~R.~M. Druffel, M.~McCarthy, L.~Gross, Direct visualization of individual
  aromatic compound structures in low molecular weight marine dissolved organic
  carbon.
\newblock {\it Geophys. Res. Lett.\/} {\bf 45}, 5590-5598 (2018).

\bibitem{Fatayer:2018by}
S.~Fatayer, N.~B. Poddar, S.~Quiroga, F.~Schulz, B.~Schuler, S.~V. Kalpathy,
  G.~Meyer, D.~P{\'e}rez, E.~Guiti{\'a}n, D.~Pe{\~n}a, M.~J. Wornat, L.~Gross,
  {Atomic Force Microscopy Identifying Fuel Pyrolysis Products and Directing
  the Synthesis of Analytical Standards}.
\newblock {\it J. Am. Chem. Soc.\/} {\bf 140}, 8156--8161 (2018).

\bibitem{Schulz2019_PROCI}
F.~Schulz, M.~Commodo, K.~Kaiser, G.~D. Falco, P.~Minutolo, G.~Meyer,
  A.~D`Anna, L.~Gross, Insights into incipient soot formation by atomic force
  microscopy.
\newblock {\it Proc. Combust. Inst.\/} {\bf 37}, 885-892 (2019).

\bibitem{Pawlak:2019eu}
R.~Pawlak, J.~G. Vilhena, A.~Hinaut, T.~Meier, T.~Glatzel, A.~Baratoff,
  E.~Gnecco, R.~Perez, E.~Meyer, {Conformations and cryo-force spectroscopy of
  spray-deposited single-strand DNA on gold}.
\newblock {\it Nat. Commun.\/} {\bf 10}, 685 (2019).

\bibitem{Albers:2009hz}
B.~J. Albers, T.~C. Schwendemann, M.~Z. Baykara, N.~Pilet, M.~Liebmann, E.~I.
  Altman, U.~D. Schwarz, {Three-dimensional imaging of short-range chemical
  forces with picometre resolution}.
\newblock {\it Nature\/} {\bf 4}, 307--310 (2009).

\bibitem{LeCun:2015dt}
Y.~LeCun, Y.~Bengio, G.~Hinton, {Deep learning}.
\newblock {\it Nature\/} {\bf 521}, 436--444 (2015).

\bibitem{Kalinin:2015cx}
S.~V. Kalinin, B.~G. Sumpter, R.~K. Archibald,
  {Big{\textendash}deep{\textendash}smart data in imaging for guiding materials
  design}.
\newblock {\it Nat. Mater.\/} {\bf 14}, 973--980 (2015).

\bibitem{Kalinin:2016dk}
S.~V. Kalinin, E.~Strelcov, A.~Belianinov, S.~Somnath, R.~K. Vasudevan, E.~J.
  Lingerfelt, R.~K. Archibald, C.~Chen, R.~Proksch, N.~Laanait, S.~Jesse, {Big,
  Deep, and Smart Data in Scanning Probe Microscopy}.
\newblock {\it ACS Nano\/} {\bf 10}, 9068--9086 (2016).

\bibitem{Ziatdinov:2017de}
M.~Ziatdinov, A.~Maksov, S.~V. Kalinin, {Learning surface molecular structures
  via machine vision}.
\newblock {\it npj Comput. Mater.\/} {\bf 3}, 31 (2017).

\bibitem{OPLS}
W.~L. Jorgensen, J.~Tirado-Rives, The opls [optimized potentials for liquid
  simulations] potential functions for proteins, energy minimizations for
  crystals of cyclic peptides and crambin.
\newblock {\it J. Am. Chem. Soc.\/} {\bf 110}, 1657-1666 (1988).

\bibitem{Moll:2010}
N.~Moll, L.~Gross, F.~Mohn, A.~Curioni, G.~Meyer, The mechanisms underlying the
  enhanced resolution of atomic force microscopy with functionalized tips.
\newblock {\it New J. Phys.\/} {\bf 12}, 125020 (2010).

\bibitem{Giessibl:2001formula}
F.~J. Giessibl, A direct method to calculate tip?sample forces from frequency
  shifts in frequency-modulation atomic force microscopy.
\newblock {\it Appl. Phys. Lett.\/} {\bf 78}, 123-125 (2001).

\bibitem{Peng:2018WeakH2O}
J.~Peng, J.~Guo, P.~Hapala, D.~Cao, R.~Ma, B.~Cheng, L.~Xu,
  M.~Ondr{\'{a}}{\v{c}}ek, P.~Jel{\'{i}}nek, E.~Wang, Y.~Jiang, {Weakly
  perturbative imaging of interfacial water with submolecular resolution by
  atomic force microscopy}.
\newblock {\it Nat. Commun.\/} {\bf 9}, 122 (2018).

\bibitem{Hapala:2016NatComm}
P.~Hapala, M.~\v{S}vec, O.~Stetsovych, N.~J. van~der Heijden,
  M.~Ondr\'a\v{c}ek, J.~van~der Lit, P.~Mutombo, I.~Swart, P.~Jel\'inek,
  {Mapping the electrostatic force field of single molecules from
  high-resolution scanning probe images}.
\newblock {\it Nat. Commun.\/} {\bf 7}, 11560 (2016).

\bibitem{Ramakrishnan:2014ij}
R.~Ramakrishnan, P.~O. Dral, M.~Rupp, O.~A. von Lilienfeld, {Quantum chemistry
  structures and properties of 134 kilo molecules}.
\newblock {\it Sci. Data\/} {\bf 1}, 201422 (2014).

\bibitem{Hapala:2014IETS}
P.~Hapala, R.~Temirov, F.~S. Tautz, P.~Jel\'{\i}nek, Origin of high-resolution
  iets-stm images of organic molecules with functionalized tips.
\newblock {\it Phys. Rev. Lett.\/} {\bf 113}, 226101 (2014).

\bibitem{Pavlicek:2012thiarene}
N.~Pavli\ifmmode~\check{c}\else \v{c}\fi{}ek, B.~Fleury, M.~Neu,
  J.~Niedenf\"uhr, C.~Herranz-Lancho, M.~Ruben, J.~Repp, Atomic force
  microscopy reveals bistable configurations of dibenzo[a,h]thianthrene and
  their interconversion pathway.
\newblock {\it Phys. Rev. Lett.\/} {\bf 108}, 086101 (2012).

\bibitem{Ellner:2016ElecCO}
M.~Ellner, N.~Pavli?ek, P.~Pou, B.~Schuler, N.~Moll, G.~Meyer, L.~Gross,
  R.~Peréz, The electric field of co tips and its relevance for atomic force
  microscopy.
\newblock {\it Nano Lett.\/} {\bf 16}, 1974-1980 (2016).

\bibitem{Ellner:2019DensOver}
M.~Ellner, P.~Pou, R.~Pérez, Molecular identification, bond order
  discrimination, and apparent intermolecular features in atomic force
  microscopy studied with a charge density based method.
\newblock {\it ACS Nano\/} {\bf 13}, 786-795 (2019).

\bibitem{Gross:2014PRB}
L.~Gross, B.~Schuler, F.~Mohn, N.~Moll, N.~Pavli\ifmmode~\check{c}\else
  \v{c}\fi{}ek, W.~Steurer, I.~Scivetti, K.~Kotsis, M.~Persson, G.~Meyer,
  Investigating atomic contrast in atomic force microscopy and kelvin probe
  force microscopy on ionic systems using functionalized tips.
\newblock {\it Phys. Rev. B\/} {\bf 90}, 155455 (2014).

\bibitem{Melitz:2011ea}
W.~Melitz, J.~Shen, A.~C. Kummel, S.~Lee, {Kelvin probe force microscopy and
  its application}.
\newblock {\it Surface Science Reports\/} {\bf 66}, 1--27 (2011).

\bibitem{sadewasser_kelvin_2012}
S.~Sadewasser, T.~Glatzel, eds., {\it {Kelvin Probe Force Microscopy: Measuring
  and Compensating Electrostatic Forces}\/} (Springer International Publishing,
  Cham, 2012).

\bibitem{Sadewasser:2018tq}
S.~Sadewasser, T.~Glatzel, eds., {\it {Kelvin Probe Force Microscopy: From
  Single Charge Detection to Device Characterization}\/} (Springer
  International Publishing, Cham, 2018).

\bibitem{Schulz:2018ck}
F.~Schulz, J.~Ritala, O.~Krej{\v c}{\'\i}, A.~P. Seitsonen, A.~S. Foster,
  P.~Liljeroth, {Elemental Identification by Combining Atomic Force Microscopy
  and Kelvin Probe Force Microscopy}.
\newblock {\it ACS Nano\/} {\bf 12}, 5274--5283 (2018).

\bibitem{Badrinarayanan:2017segnet}
V.~{Badrinarayanan}, A.~{Kendall}, R.~{Cipolla}, Segnet: A deep convolutional
  encoder-decoder architecture for image segmentation.
\newblock {\it IEEE Transactions on Pattern Analysis and Machine
  Intelligence\/} {\bf 39}, 2481-2495 (2017).

\bibitem{Glorot:2011a}
X.~Glorot, A.~Bordes, Y.~Bengio, {\it Proceedings of the Fourteenth
  International Conference on Artificial Intelligence and Statistics\/} (PMLR,
  2011), vol.~15 of {\it Proceedings of Machine Learning Research\/}, pp.
  315--323.

\bibitem{Chollet:2015keras}
F.~Chollet, {\it et~al.\/}, Keras, \url{https://keras.io} (2015).

\bibitem{Abadi:2015tf}
M.~Abadi, A.~Agarwal, P.~Barham, E.~Brevdo, Z.~Chen, C.~Citro, G.~S. Corrado,
  A.~Davis, J.~Dean, M.~Devin, S.~Ghemawat, I.~Goodfellow, A.~Harp, G.~Irving,
  M.~Isard, Y.~Jia, R.~Jozefowicz, L.~Kaiser, M.~Kudlur, J.~Levenberg,
  D.~Man\'{e}, R.~Monga, S.~Moore, D.~Murray, C.~Olah, M.~Schuster, J.~Shlens,
  B.~Steiner, I.~Sutskever, K.~Talwar, P.~Tucker, V.~Vanhoucke, V.~Vasudevan,
  F.~Vi\'{e}gas, O.~Vinyals, P.~Warden, M.~Wattenberg, M.~Wicke, Y.~Yu,
  X.~Zheng, {TensorFlow}: Large-scale machine learning on heterogeneous systems
  (2015). Software available from tensorflow.org.

\bibitem{Devries2017Cutout}
T.~Devries, G.~W. Taylor, Improved regularization of convolutional neural
  networks with cutout.
\newblock {\it CoRR\/} {\bf abs/1708.04552} (2017).

\bibitem{Dagdeviren:2018jm}
O.~E. Dagdeviren, C.~Zhou, E.~I. Altman, U.~D. Schwarz, {Quantifying Tip-Sample
  Interactions in Vacuum Using Cantilever-Based Sensors: An Analysis}.
\newblock {\it Phys. Rev. Appl.\/} {\bf 9}, 044040 (2018).

\bibitem{Sader:2018ey}
J.~E. Sader, B.~D. Hughes, F.~Huber, F.~J. Giessibl, {Interatomic force laws
  that evade dynamic measurement}.
\newblock {\it Nat. Nanotechnol.\/} {\bf 13}, 1088--1091 (2018).

\bibitem{Becke:1993is}
A.~D. Becke, {Density-functional thermochemistry. III. The role of exact
  exchange}.
\newblock {\it J. Chem. Phys.\/} {\bf 98}, 5648--5652 (1993).

\bibitem{Parrish:2017:psi4}
R.~M. Parrish, L.~A. Burns, D.~G.~A. Smith, A.~C. Simmonett, A.~E. DePrince,
  E.~G. Hohenstein, U.~Bozkaya, A.~Y. Sokolov, R.~Di~Remigio, R.~M. Richard,
  J.~F. Gonthier, A.~M. James, H.~R. McAlexander, A.~Kumar, M.~Saitow, X.~Wang,
  B.~P. Pritchard, P.~Verma, H.~F. Schaefer, K.~Patkowski, R.~A. King, E.~F.
  Valeev, F.~A. Evangelista, J.~M. Turney, T.~D. Crawford, C.~D. Sherrill, Psi4
  1.1: An open-source electronic structure program emphasizing automation,
  advanced libraries, and interoperability.
\newblock {\it J. Chem. Theory Comput.\/} {\bf 13}, 3185-3197 (2017).

\bibitem{Crawford:2007:cc}
T.~D. Crawford, H.~F. Schaefer~III, {\it An Introduction to Coupled Cluster
  Theory for Computational Chemists\/} (John Wiley \& Sons, Ltd, 2007), pp.
  33--136.

\bibitem{Bartels:1997}
L.~Bartels, G.~Meyer, K.-H. Rieder, Controlled vertical manipulation of single
  co molecules with the scanning tunneling microscope: A route to chemical
  contrast.
\newblock {\it Appl. Phys. Lett.\/} {\bf 71}, 213-215 (1997).

\end{thebibliography}
\end{document}

	
	\baselineskip24pt
	
	
	\maketitle 
	
	\newpage
	
\tableofcontents

\newpage

\section{Image representations of output molecular structure}

The chosen image representation is even more important when considering non-ideal experimental conditions and the inherent approximations present in any simulation model generating the training examples. The information that we try to reconstruct should be not only present in AFM data, but it must also be robust with respect to minor errors and unknowns. In particular, the determination of atoms which are too deep and therefore do not significantly contribute to the AFM signal is an ill-posed problem, and we should therefore avoid output representations that pretend to provide information about such atoms.

To aid in validation, we should also output a representation which is convenient for perception of the human user as well as for design of the neural network architecture. Representation of the molecular structure as a list of atomic coordinates is generally considered problematic for machine learning, as the size of such an output vector would differ for each molecule, and it does not respect permutation symmetry (i.e. exchange of atoms corresponding to the same element in the list produces apparently different descriptors, despite representing physically identical structures). Various methods were developed in the past to overcome these problems (e.g Atom-centered symmetry functions \cite{Behler:2007HDGNN}, or smooth overlap of atomic positions \cite{Bartok:2013SOAP}), but do not present molecular geometry in a human-readable form. Image-like representations are an intuitive choice for humans, projecting the system geometry into a scalar or complex field sampled on a regular real-space grid \cite{Kuzminykh:2018WaveRepresentation}.

The output property has to be generated together with simulated AFM data during training, therefore the definition of the output property is closely related to the algorithm by which it is generated. In this work, we designed three distinct image descriptors representing 2D image projections of molecular geometries (see \reffig{figRepresentations}).

\paragraph{Height Map}

For each pixel of the image, we calculated the depth at which the vertical component of forces between sample and tip apex becomes more repulsive than some constant value (typically $\approx 0.1 eV/$\AA ). The resulting image (see \reffig{figRepresentations}e,h) should roughly correspond to a hypothetical AFM image obtained in constant-force mode with a non-flexible tip (unlike CO). It also roughly corresponds to the concept of the solvent-accessible surface introduced in biochemistry \cite{Connolly:1983SAS}, indicating areas accessible to the probe particle. This is also very useful to rationalize the formation of the imaging contrast and supramolecular interactions in general. In the future we plan also to map the electrostatic potential on top of this surface. 

\paragraph{van der Waals Spheres}

While the Height Map corresponds to an isosurface of aggregate force (with contributions from all atoms), a representation of molecular structure by van der Waals spheres is commonly used in chemical visualization programs (e.g. Jmol), as it is intuitive for the human user. This also follows naturally from the LJ-potential based PP method used in simulations. The \vdWSph~representation (see \reffig{figRepresentations}f,i) shows creases between atoms, in contrast to the blunt shapes of the Height Map. This was one motivation to introduce this output descriptor - it allows us to see how well discrete atoms can be recognized from the images. In the nomenclature of neural networks, we can roughly relate the Height Map to a soft-max-operation and \vdWSph to simple max-operation over the contributions of atoms to the force field.  

By its nature, the van der Waals Spheres representation encodes the depth of an atom's position by brightness of sphere and atomic radius (which is connected with atomic type) by sphere size. Therefore we considered modification of the CNN learning process for the \vdWSph~representation such that different sphere radii would be split by colour for more convenient atomic type identification - we call this modification \emph{type map}. In its current form, it allows identification of 3 different categories of atoms according to sphere radii: hydrogen (red), non-hydrogen peripheral (green) and carbon backbone (blue). 

\paragraph{Atomic Disks}

This descriptor goes one step further from a quantity directly related to the force field towards a more abstract representation. Here we render small disks (with the brightness conically decreasing from the center) onto positions of atomic nuclei (see \reffig{figRepresentations}g,j). The brightness indicates the height of atoms, while the size of the disk is proportional to the covalent radius of atom.

\subsection{Comparison of the different output representations of molecular structure generated from simulated AFM input}

\begin{figure*}[h]
\centering
\includegraphics[width=80mm]{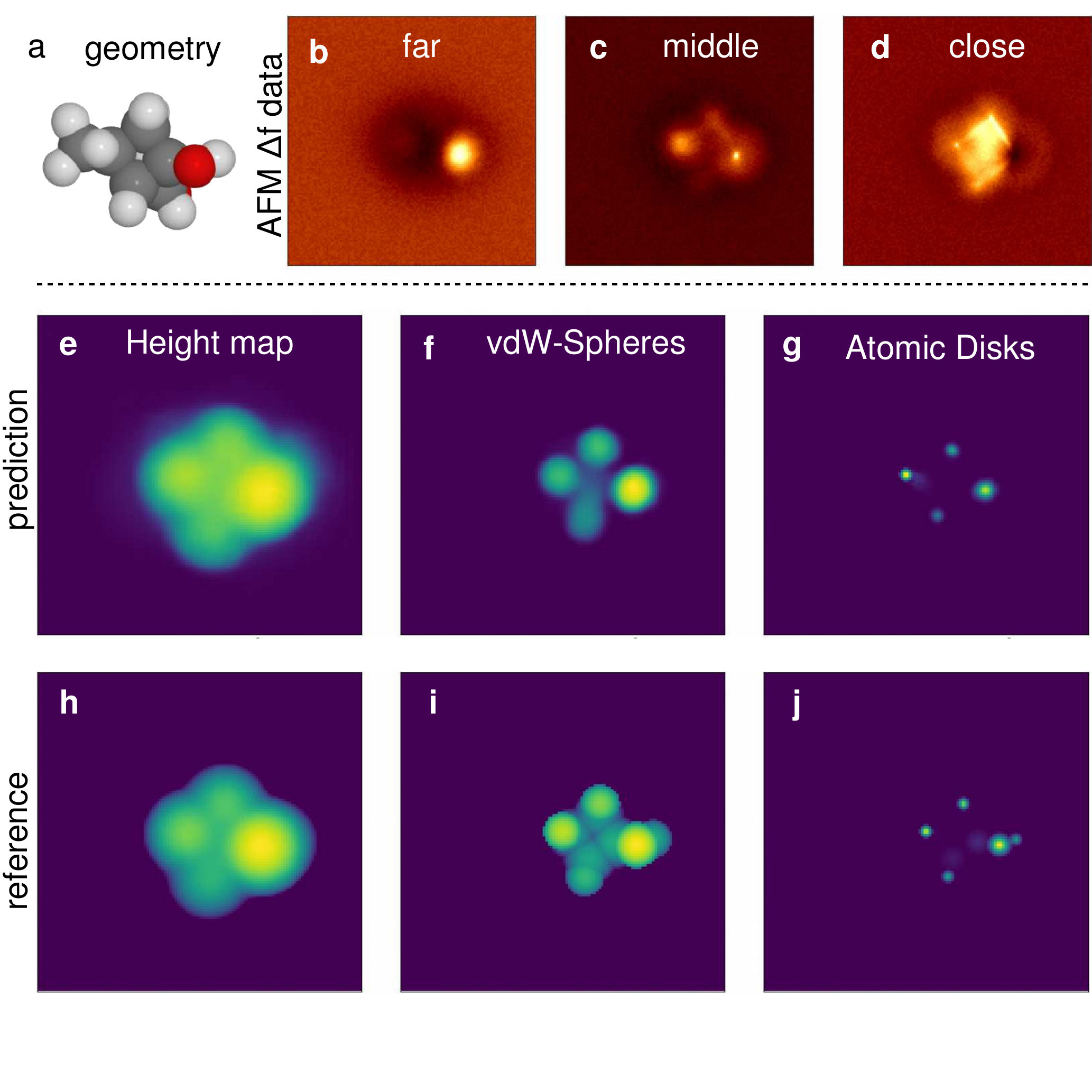}
\caption{ { \bf Different 2D image representations of the output geometry $X$ for simulated AFM images of a C$_7$H$_{10}$O$_2$ molecule from the training set.}  \textbf{a} 3D render of molecular geometry; \textbf{b-d} simulated AFM $\Delta f$ images with decreasing tip-sample distance; \textbf{e-g} 2D Image representation predicted by our CNN; \textbf{h-j} Reference 2D Image representation produced directly from the molecule geometry.
}
\label{figRepresentations}
\end{figure*}

\begin{figure*}[h]
\centering
\includegraphics[width=80mm]{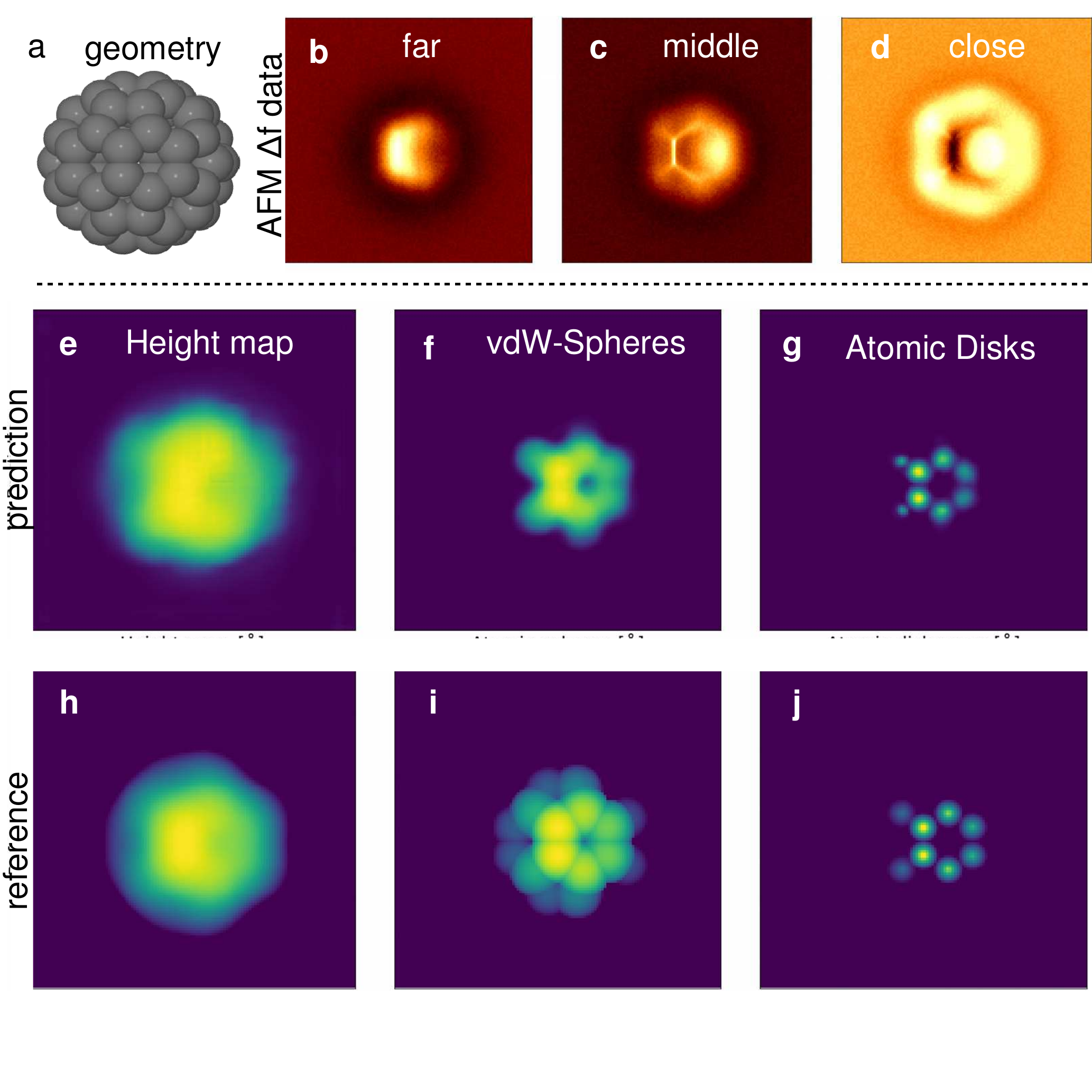}
\caption{ { \bf Different 2D image representations of the output geometry $X$ for simulated AFM images of a $C_{60}$ molecule.}  \textbf{a} 3D render of molecular geometry; \textbf{b-d} simulated AFM $\Delta f$ images with decreasing tip-sample distance; \textbf{e-g} 2D Image representation predicted by our CNN; \textbf{h-j} Reference 2D Image representation produced directly from the molecule geometry.
}
\label{figRepresentations_C60}
\end{figure*}

\begin{figure*}[h]
\centering
\includegraphics[width=80mm]{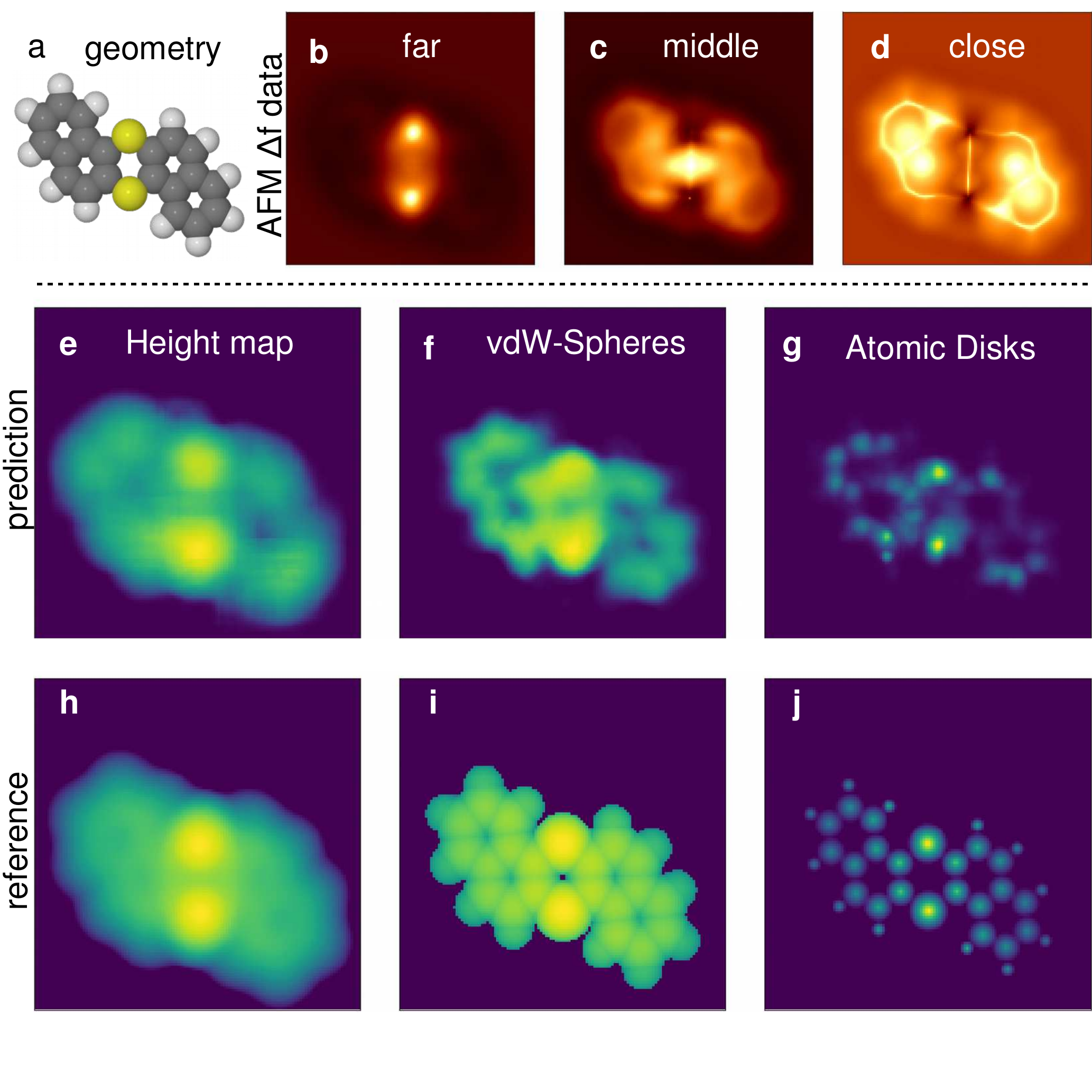}
\caption{ { \bf Different 2D image representations of the output geometry $X$ for simulated AFM images of Dibenzo[a,h]thianthrene molecule \cite{Pavlicek:2012thiarene}.}  \textbf{a} 3D render of molecular geometry; \textbf{b-d} simulated AFM $\Delta f$ images with decreasing tip-sample distance; \textbf{e-g} 2D Image representation predicted by our CNN; \textbf{h-j} Reference 2D Image representation produced directly from the molecule geometry. Note that the Atomic Disk representation fails to reliably predict the exact positions for atoms of the aromatic systems near to the sulphurs, but located \emph{deeper} i.e. lower in contrast. Failure to predict deeper atoms is a typical problem of our NN. It follows up from the "conservatism" enforced by regularization techniques (e.g. related to noise and dropouts). If the CNN does not accumulate enough evidence, it omits atoms completely instead of predicting them at wrong positions. In this particular case, the prediction of lower lying atoms is perhaps disturbed by the dominant sulphur atoms.
}
\label{figRepresentations_thiarene}
\end{figure*}

Here (\reffig{figRepresentations} -- \reffig{figRepresentations_thiarene}) we compare predictions of other representations for molecules from Fig.~2 of the main text. In general it can be said that decreasing the size of atomic features (from Height Map to Atomic Disks) helps to better recognize the atoms by a human user, but also it makes the training more demanding and output less robust. Hence, the selection remains a compromise, and we are planning to implement simultaneous calculation of all descriptors (these and other descriptors in development) during the simulation process.

\FloatBarrier
\subsection{Examples of molecules from the training set}

\begin{figure*}[!!!!ht]
\centering
\includegraphics[width=160mm]{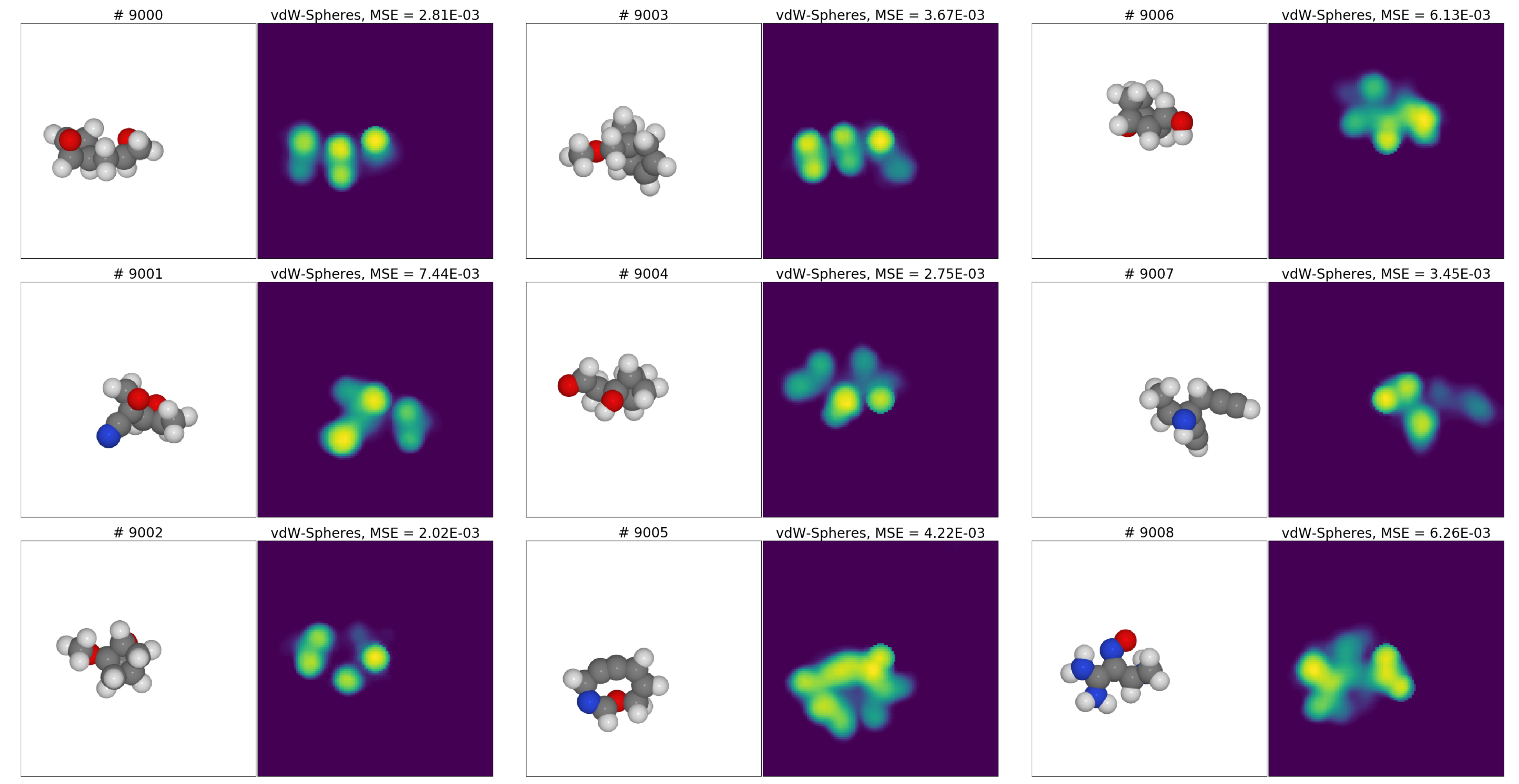}
\caption{ { \bf Molecules from the validation data set together with the \vdWSph~representation predicted by the CNN} Here are presented 9 simulated random molecules from our validation data set. Each pair of images contains: {\it left} -- molecular geometry where color represents the type of atoms (white - {\bf H}, gray - {\bf C}, blue - {\bf N}, red - {\bf O}); {\it right} -- \vdWSph~representation of the structure predicted by the CNN. The orientation of the molecule is selected to maximize the number of visible atoms, as described in the main text. }  
\label{SI_figValidSampl}
\end{figure*}

\FloatBarrier
\section{Matching experiment to relaxed on-surface simulated configurations}

From a physical point of view, it is reasonable to match experimental AFM images only to those simulated orientations of a molecule which represent some local minimum on the surface. Nevertheless, such an approach is strongly dependent on the ability to reliably find all such local minima (possible configurations) by purely computational means, which is still generally an unsolved task. Although the development of novel global configuration search methodology is beyond the scope of this paper, we tentatively examined this idea. As an initial approximation, we used rigid-body molecular mechanics \cite{Hapala:2018RigidMol} with only a van der Waals force field to relax 500 uniformly distributed initial rotations of the camphor molecule into 7 distinct local minima. Then we further relaxed those configurations with the density functional package cp2k \cite{Hutter:2014cp2k}. These local energy minimum configurations were then compared to a set of experimental configurations using a linear correlation metric.  

\FloatBarrier
\begin{figure*}[h]
\centering
\includegraphics[width=160mm]{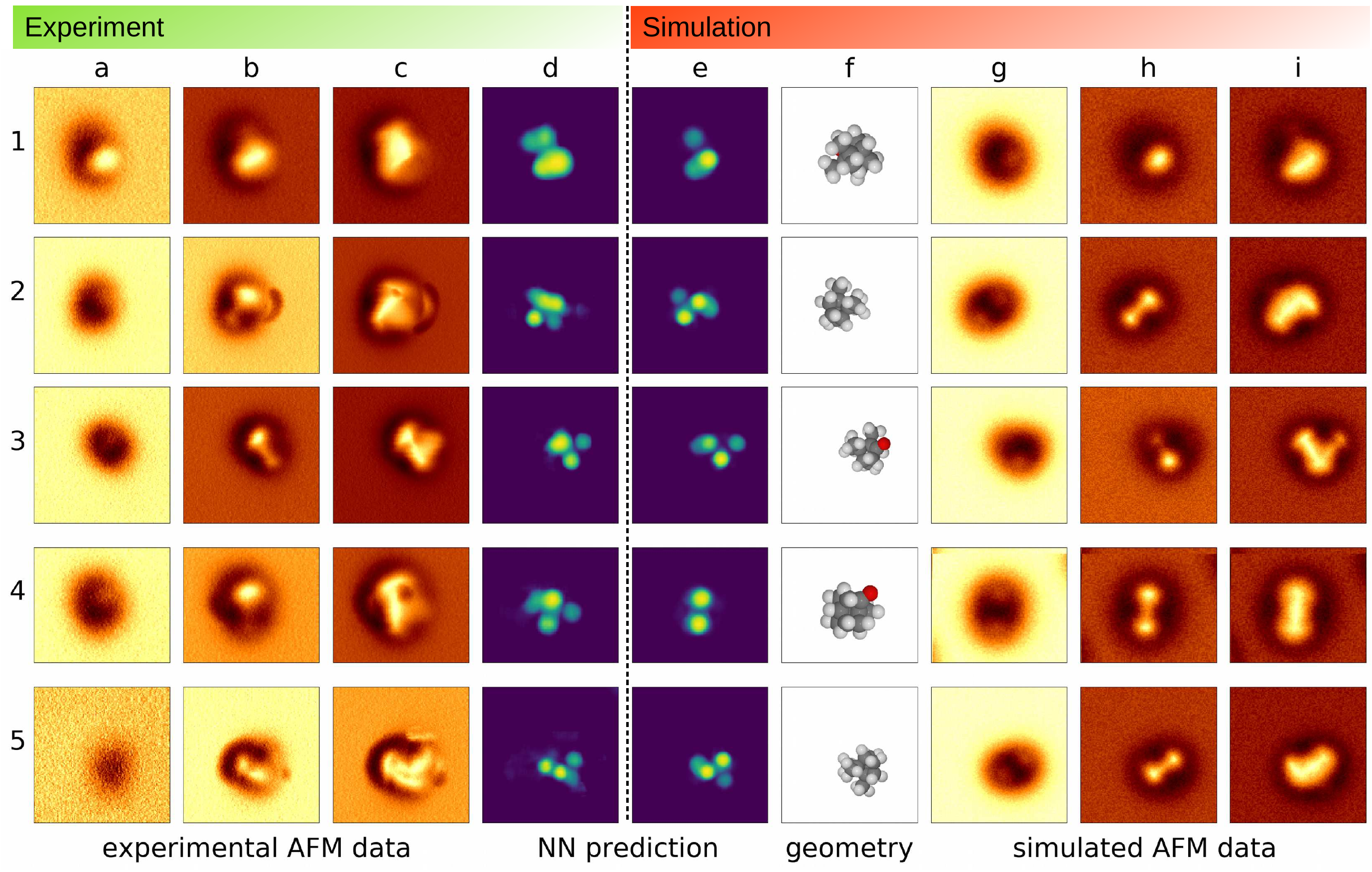}
\caption{ { \bf Matching between simulated relaxed configurations of 1S-Camphor and experiment.} \textbf{1-5} refer to distinct molecular configurations with experiments in columns \textbf{a-d} and simulations in columns \textbf{e-i}. Selected experimental AFM images (out of 10 slices used for input): \textbf{a} -- far, \textbf{b} -- middle, \textbf{c} -- close tip-sample distances and NN prediction \textbf{d} for the \vdWSph  representation. The \vdWSph  representation \textbf{e} corresponds to the full molecular configuration \textbf{f} resulting in the best match with the experiments. The corresponding selected simulated AFM images are given in panels \textbf{g-i}. }
\label{figExpRelaxed}
\end{figure*}
\FloatBarrier

The CNN model's predicted image representations for the molecular structure (\vdWSph) on \reffig{figExpRelaxed} allow for the assignment of simulated configurations with known geometry (column (f)) for each experimental configuration. For the experiments considered in this work, we have only a few experimental and simulated configurations, allowing for easy validation by a human expert. We did not find better agreement between corresponding experimental and simulated configurations. The differences which we could see between experiment and simulation in this case are not connected with predictions of the CNN model and are more likely related to the accuracy of computationally relaxed configurations. Nevertheless, in the next sections we consider possible small adjustments to the accuracy of matching.

\FloatBarrier
\section{Effect of small perturbations on AFM imaging and matching}

To take account of possible tip- and surface-induced perturbations in the imaging process, we tried to address possible variations of molecular configuration for the case of our benchmark molecule 1S-camphor in two ways: 1) Tilting of the whole molecule as a rigid body and 2) rotation of $-CH_3$ groups. We found that both these perturbations quite significantly affect the simulated AFM images and associated configuration predictions.

\FloatBarrier
\subsection{Effect of molecule tilting}

\begin{figure*}[h]
\centering
\includegraphics[width=\linewidth]{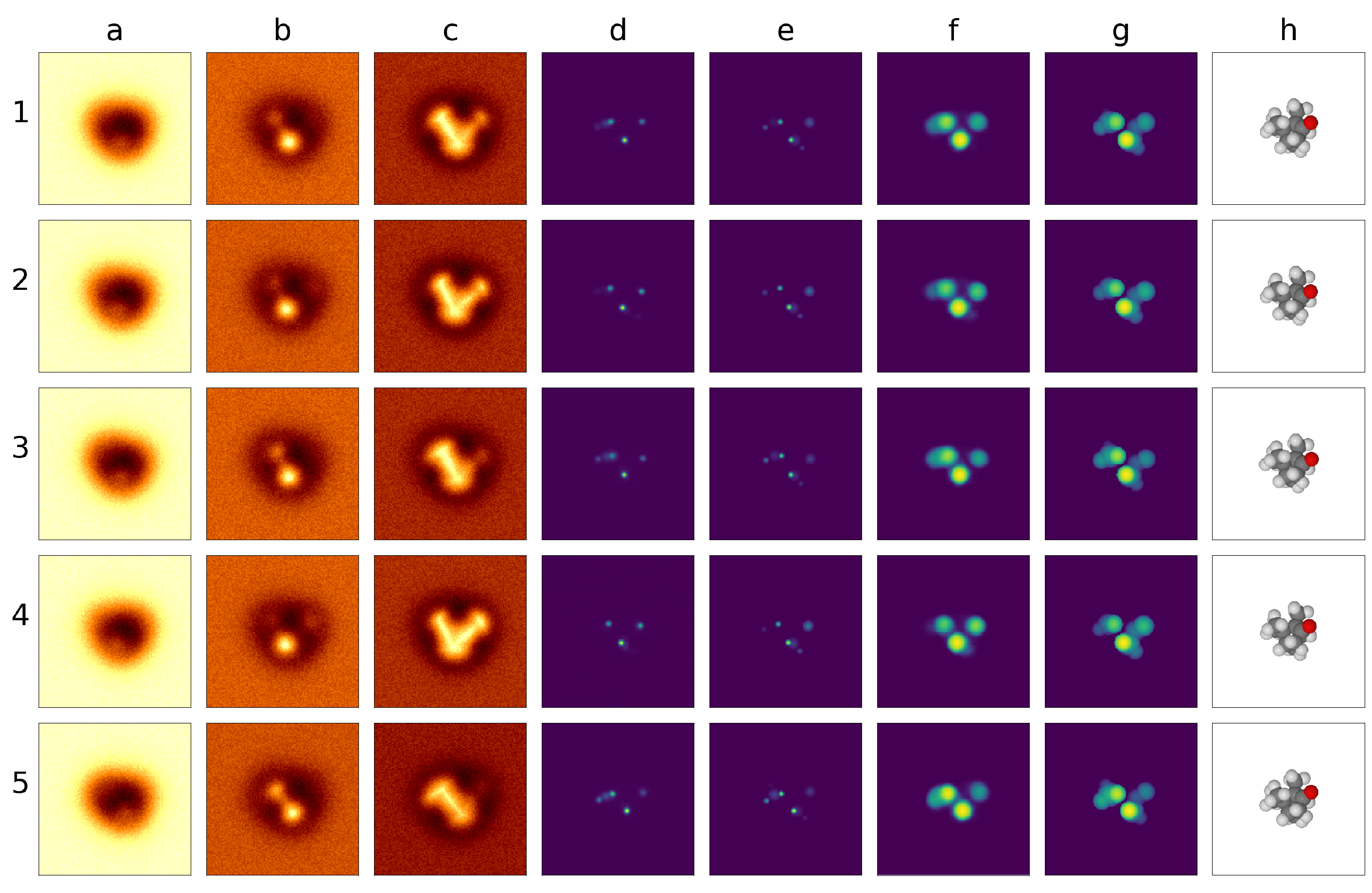}
\caption{ { \bf Effect of tilt of molecules on simulated AFM images} \textbf{1-5} Simulated configuration tilted with a small random angle from the normal to the Cu(111) surface; \textbf{a-c} 3 selected simulated AFM images: \ \textbf{a} -- far, \textbf{b} -- middle, \textbf{c} -- close tip-sample distance; \textbf{d} - atomic disks representation as predicted by the CNN from a given set of simulated AFM images; \textbf{e} - atomic disks representation as a reference; \textbf{f} - vdW-spheres representation as predicted by the CNN; \textbf{g} - vdW-spheres representation as a reference. \textbf{h}~Full molecular geometry corresponding to simulated AFM images.  }
\label{figTiltVar}
\end{figure*}

At first we consider angular tilting of the molecule from  model configurations by 0-5$^{\circ}$ from normal to the Cu(111) surface. The resultant images are presented in \reffig{figTiltVar} -- there are 5 different tilt variations of the same simulated molecule configuration. Even such small angular tilting of the molecule from normal to the surface significantly affects the AFM images. This follows from the nature of the AFM imaging process, where the tip "scratches" the surface of the molecule and upper atoms dominate the interaction. Even a small height change of the top-most atoms, such as 50 pm, can have visible impact on the simulated AFM image. Image descriptors of simulated configuration also reflect these variations to a degree, but they are not as obvious as in the AFM data. For close configurations, the image descriptors are also similar and their features even allow estimates for the direction of molecular tilting. This feature emphasizes that image descriptors are a more suitable representation for machine learning than raw AFM images.

\FloatBarrier
\subsection{Effect of -CH$_3$ rotations}

\begin{figure*}[h]
\centering
\includegraphics[width=150mm]{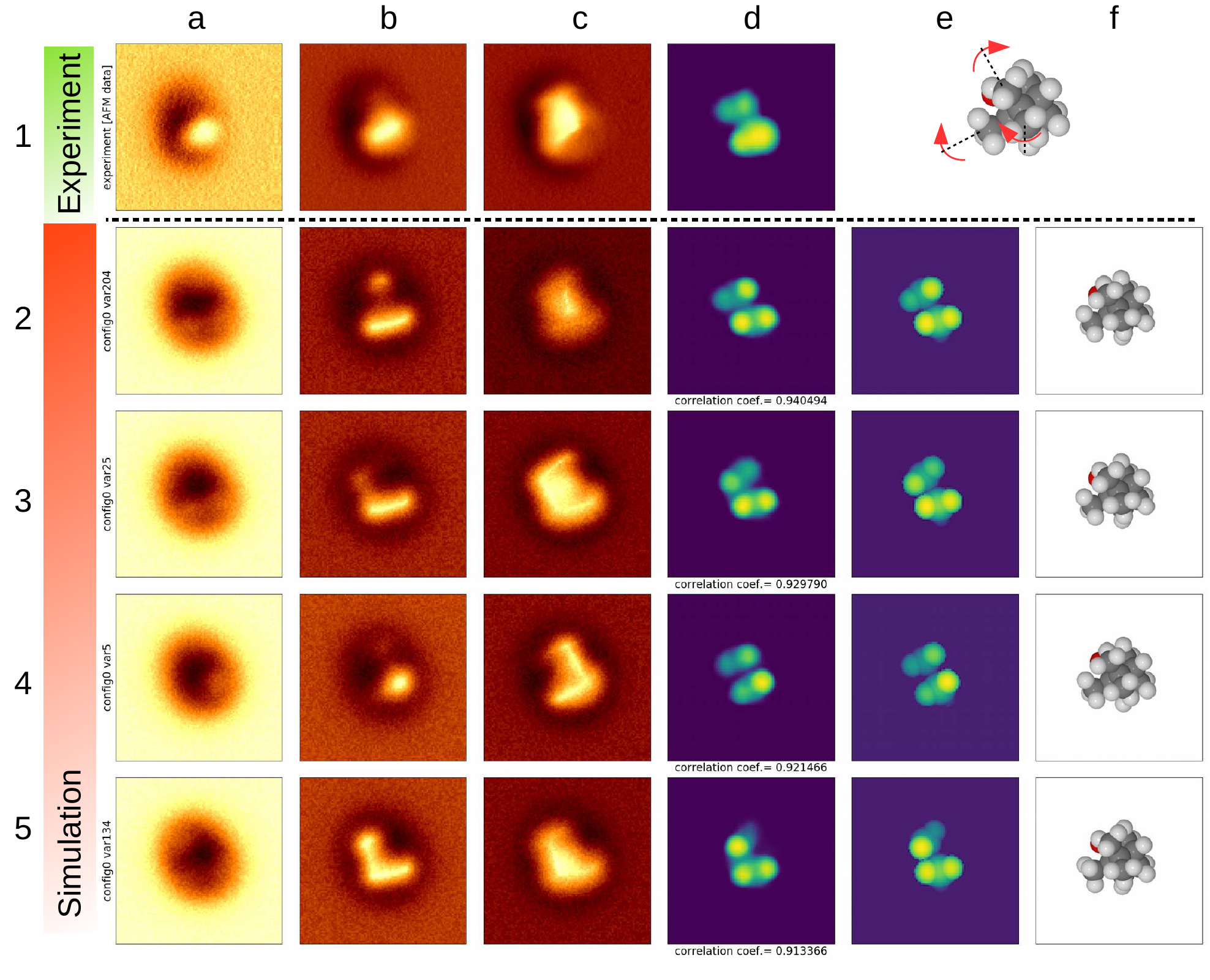}
\caption{ { \bf Adjustment of simulated configuration by -CH$_3$ group rotations.} \textbf{1} - Experimental configuration \# 1; \textbf{2-5} - Simulated configuration with different rotations of the -CH$_3$ groups. \textbf{a-c} selected AFM images: \textbf{a} -- far, \textbf{b} -- middle, \textbf{c} -- close tip-sample distance; \textbf{d} - \vdWSph~representation as predicted by the CNN  and \textbf{e} - references. \textbf{f}~Full molecular geometry corresponding to simulated AFM data.
}
\label{figCH3rot}
\end{figure*}

For simplicity we assumed that the geometry of the molecule is rigid, i.e. that internal degrees of freedom are frozen. While this is typically true for flat aromatic molecules, in the case of 3D aliphatic molecules, the barriers for rotation around a single-bond between two carbon atoms is of the order of just ~10 kJ/mol, therefore it can be induced during deposition or scanning. Hence, we considered the possible impact of such rotations in the case of 1S-Camphor. In this molecule there are 3 -CH$_3$ groups capable of such rotations. AFM images are very sensitive to changes of atomic positions of the closest atoms. Therefore, despite not affecting the global minimum energy configuration of the molecule, the effects of rotating a -CH$_3$ can be crucial. Examples of changes are presented in \reffig{figCH3rot}, where we considered one simulated Camphor configuration. This configuration has 3 different -CH$_3$ groups sticking up (see upper right corner on \reffig{figCH3rot}) that could be rotated independently, and the result of the rotations on images and descriptors is shown in rows (1-5).

\FloatBarrier
\subsection{Ambiguity of molecular orientation}

\begin{figure*}[t]
\centering
\includegraphics[width=\linewidth]{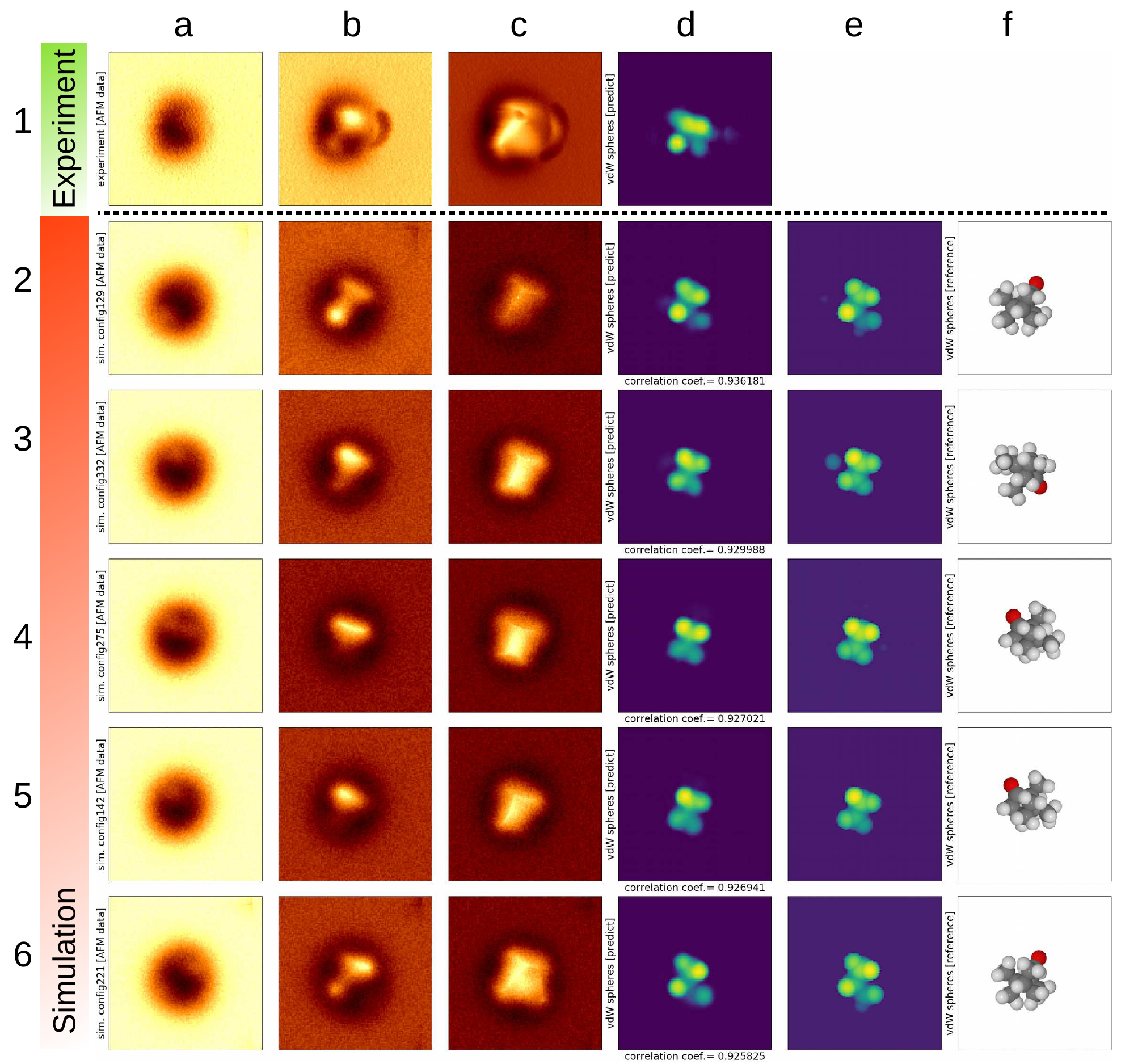}
\caption{ { \bf Matching experimental configuration \textbf{2} of 1S-Camphor with closest simulated configurations.} 1 - Experimental configuration \# 3; \textbf{2-6} Distinct simulated configurations ordered by descending similarity of \vdWSph; \textbf{a-c} selected AFM images: \textbf{a} -- far, \textbf{b} -- middle, \textbf{c} -- close tip-sample distance; \textbf{d} -- \vdWSph~representation as predicted by the CNN  and \textbf{e} references. \textbf{f}~Full molecular geometry corresponding to simulated AFM images. }
\label{figExpVar}
\end{figure*}

For a 3D molecule such as 1S-Camphor, the AFM tip is able to directly probe just 2-3 surface atoms. This limited information is not sufficient to always reliably discriminate a single orientation of the molecule. In particular, it may easily happen that a similar doublet or triplet of atoms is present on several places of the molecular surface. In \reffig{figExpVar} we present 5 best matches from 500 uniformly distributed rotations to a single experimental stack. We found that two very distinct configurations (row 3,4,5 vs. 2,6) both match experiment with a linear correlation coefficient of the \vdWSph $>$0.92. Note that configuration in row 3 is close to a symmetric rotation of 180 degrees of the configurations in rows 4 and 5. For more specific molecular identification, either further experimental configurations are required (as in this work) or further refinement requires input from multiple descriptors in parallel. However, even if this was the only data available, the possible configurations have still been greatly reduced with very little computational effort.

\FloatBarrier
\section{Neural network architecture}
\begin{figure}[!b]
	\tikzset{
	encoder/.pic={
		\path (0,0,0) pic [fill=green, opacity=0.4] {annotated cuboid={width=0.6, height=2.5, depth=2.5}}
		++(0.6,-0.25,-0.25) pic [fill=blue, opacity=0.65] {annotated cuboid={width=0.3, height=2.0, depth=2.0}}
		++(1.3,0,0) pic [fill=green, opacity=0.4] {annotated cuboid={width=0.3, height=2.0, depth=2.0}}
		++(0.4,-0.25,-0.25) pic [fill=blue, opacity=0.65] {annotated cuboid={width=0.15, height=1.5, depth=1.5}}
		++(0.95,0,0) pic [fill=green, opacity=0.4] {annotated cuboid={width=0.15, height=1.5, depth=1.5}}
		++(0.3,-0.25,-0.25) pic [fill=blue, opacity=0.65] {annotated cuboid={width=0.15, height=1.0, depth=1.0}};
	},
	% %
	decoder/.pic={
		\path (0,0.25,0.25) pic [fill=yellow, opacity=0.6] {annotated cuboid={width=0, height=1.5, depth=1.5}}
		++(0.3,0,0) pic [fill=cyan, opacity=0.5] {annotated cuboid={width=0, height=1.5, depth=1.5}}
		++(0.3,0,0) pic [fill=cyan, opacity=0.5] {annotated cuboid={width=0, height=1.5, depth=1.5}}
		++(1.0,0.25,0.25) pic [fill=yellow, opacity=0.6] {annotated cuboid={width=0, height=2.0, depth=2.0}}
		++(0.3,0,0) pic [fill=cyan, opacity=0.5] {annotated cuboid={width=0, height=2.0, depth=2.0}} 
		++(0.3,0,0) pic [fill=cyan, opacity=0.5] {annotated cuboid={width=0, height=2.0, depth=2.0}} 
		++(1.2,0.25,0.25) pic [fill=yellow, opacity=0.6] {annotated cuboid={width=0, height=2.5, depth=2.5}}
		++(0.3,0,0) pic [fill=cyan, opacity=0.5] {annotated cuboid={width=0, height=2.5, depth=2.5}} 
		++(0.3,0,0) pic [fill=cyan, opacity=0.5] {annotated cuboid={width=0, height=2.5, depth=2.5}};
	}
	}
	\centering
	\begin{tikzpicture}
		
		%-DSH
		\path (0,0) pic {encoder}
		
		++(4.3,-0.75,-0.75) pic [fill=cyan, opacity=0.5] {annotated cuboid={width=0, height=1.0, depth=1.0}}
		++(0.3,0,0) pic (a) [fill=cyan, opacity=0.5] {annotated cuboid={width=0, height=1.0, depth=1.0}};
		
		\draw[-{Latex[width=0.2cm, length=0.3cm]}] ($(a)+(0.3,-0.5,-0.5)$) -- ++(1.45, 0);
		\draw[-{Latex[width=0.2cm, length=0.3cm]}] ($(a)+(1.15,-0.5-2.8,-0.5)$) -- ++(0.6, 0);
		\draw[-{Latex[width=0.2cm, length=0.3cm]}] ($(a)+(1.15,-0.5+2.8,-0.5)$) -- ++(0.6, 0);
		\draw ($(a)+(1.15,-0.5-2.8,-0.5)$) -- ++(0, 5.6);
		
		\path ($(a) + (1.85,2.8)$) pic {decoder}
		++(0,-2.8) pic {decoder}
		++(0,-2.8) pic {decoder}
		++(5,0.5+5.6,0) pic [fill=magenta, opacity=0.5] {annotated cuboid={width=0, height=2.5, depth=2.5}}
		++(0,-2.8,0) pic [fill=blue, opacity=0.5] {annotated cuboid={width=0, height=2.5, depth=2.5}}
		++(0,-2.8,0) pic [fill=blue, opacity=0.5] {annotated cuboid={width=0, height=2.5, depth=2.5}};

		\path (0,-3) node (rect) [draw, minimum width=0.1cm, minimum height=0.1cm, fill=green, opacity=0.4, label=east:{\footnotesize 3D Conv + LeakyReLU}] {}
		++(0,-0.4) node (rect) [draw, minimum width=0.1cm, minimum height=0.1cm, fill=blue, opacity=0.65, label=east:{\footnotesize Avg Pool}] {}
		++(0,-0.4) node (rect) [draw, minimum width=0.1cm, minimum height=0.1cm, fill=magenta, opacity=0.5, label=east:{\footnotesize 2D Conv}] {}
		++(0,-0.4) node (rect) [draw, minimum width=0.1cm, minimum height=0.1cm, fill=cyan, opacity=0.5, label=east:{\footnotesize 2D Conv + LeakyReLU}] {}
		++(0,-0.4) node (rect) [draw, minimum width=0.1cm, minimum height=0.1cm, fill=blue, opacity=0.5, label=east:{\footnotesize 2D Conv + ReLU}] {}
		++(0,-0.4) node (rect) [draw, minimum width=0.1cm, minimum height=0.1cm, fill=yellow, opacity=0.6, label=-0:{\footnotesize NN upsample}] {};
	
	\end{tikzpicture}
    \caption{ {\bf Illustration of the layers of the model}. The forward direction is from left to right. The sizes of the layers represent the relative size of the feature maps. Not to scale.}
    \label{SI_NNcheme}
\end{figure}
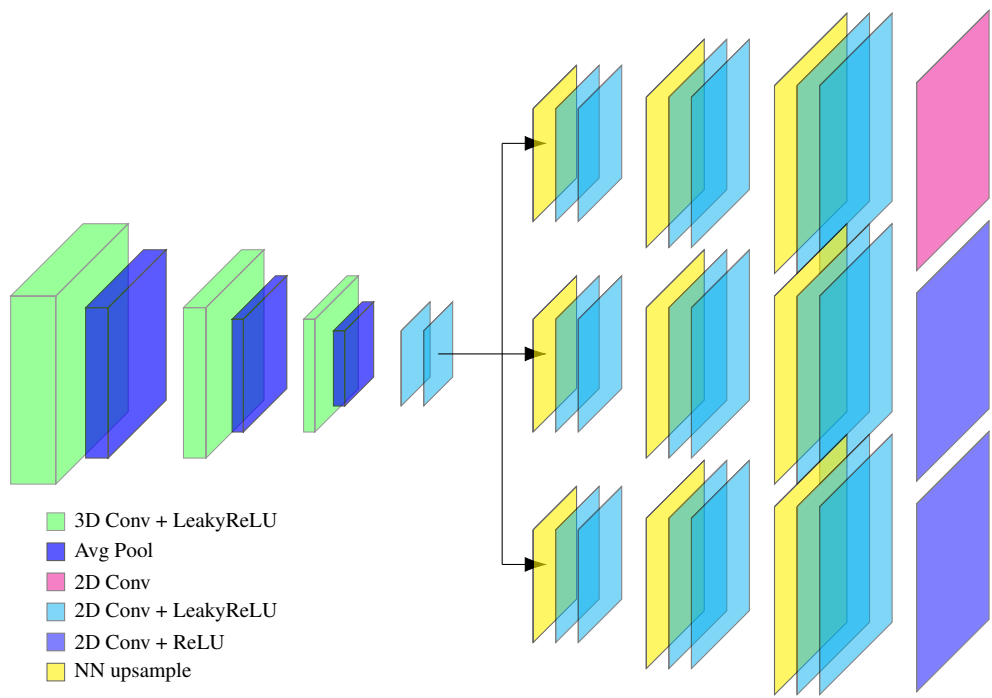

To predict 2D image representations from a stack of AFM images we used a convolutional neural network (CNN) approach. Our model is implemented in Keras \cite{Chollet:2015keras} with a Tensorflow \cite{Abadi:2015tf} backend. The network is similar to encoder-decoder type networks \cite{Badrinarayanan:2017segnet} and equipped with 12 convolutional layers. The input AFM data size was (128x128x10) and the output image representation was (128x128). The range in z is limited by the induced instabilities at close approach and the usefulness of contrast features at longer distance, and we found that we have a “useful” actual distance range of 1-1.5 \AA{} and the contrast evolves rather smoothly over this range in 10 steps. A schematic view of the network is presented in \reffig{SI_NNcheme}. Initially the AFM data stack contains features that have lateral (x,y) and depth (z) dependencies, so we use 3D convolutional layers for data processing ('3D Conv' on \reffig{SI_NNcheme}). The model has three 3D convolution layers each followed by an average pooling layer that reduces the size of the feature maps by a factor of 2 in each dimension, except for the last pooling layer which does not reduce the size in the z-direction. The motivation for this is to avoid reducing the size in the z-direction too much before moving to 2D feature maps. We also tried max pooling layers, but found average pooling to work better, possibly due to max pooling losing too much information. The transition to 2D happens by reshaping the 3D image slices into channels of a 2D feature map. Following this, a pair of 3x3 2D convolutions were applied.

The feature maps are upsampled in three stages, as in the downsampling. The upsampling is done by simply using nearest neighbour interpolation followed by two 2D convolutions. The final convolution serves to reduce the number of feature maps to one. In earlier attempts we tried using transposed convolutions for the upsampling, but found that output density maps contained patterned artifacts. This has been known to be a problem with transposed convolutions\cite{odena2016deconvolution}. Another related method is the subpixel convolution \cite{shi2016deconvolution}, which could mitigate the effect of the artifacts \cite{aitken2017checkerboard}.

All the convolutional layers use a $\mathrm{LeakyReLU}_{0.1}$ activation, except the last one which has a $\mathrm{ReLU}$ activation. We tried using $\mathrm{ReLU}$ in all layers, but this resulted in "dead" filters that gave zero or nearly zero output everywhere, and we found that $\mathrm{LeakyReLU}$ yielded better results in practice. Using regular $\mathrm{ReLU}$ for the last layer is a natural choice since the output has a clear zero point at the cut-off height that we choose. The $\mathrm{ReLU}$ and $\mathrm{LeakyReLU}$ functions are defined as:

\begin{align}
	\mathrm{ReLU}(z) &= \max(0,z) \\
	\mathrm{LeakyReLU}_\alpha(z) &=
	\begin{cases}
		z &\text{if } z \geq 0 \\
		\alpha z &\text{if } z < 0.
	\end{cases}
\end{align}

The loss function was the mean squared error. The optimizer for the gradient descent was the Adaptive Moment Estimation (Adam)\cite{kingma2014adam} optimizer. We set the learning rate to 0.001 and the decay to 10$^{-5}$, and otherwise we use the default parameters as defined in Keras.

\begin{figure*}[t]
\centering
\includegraphics[width=80mm]{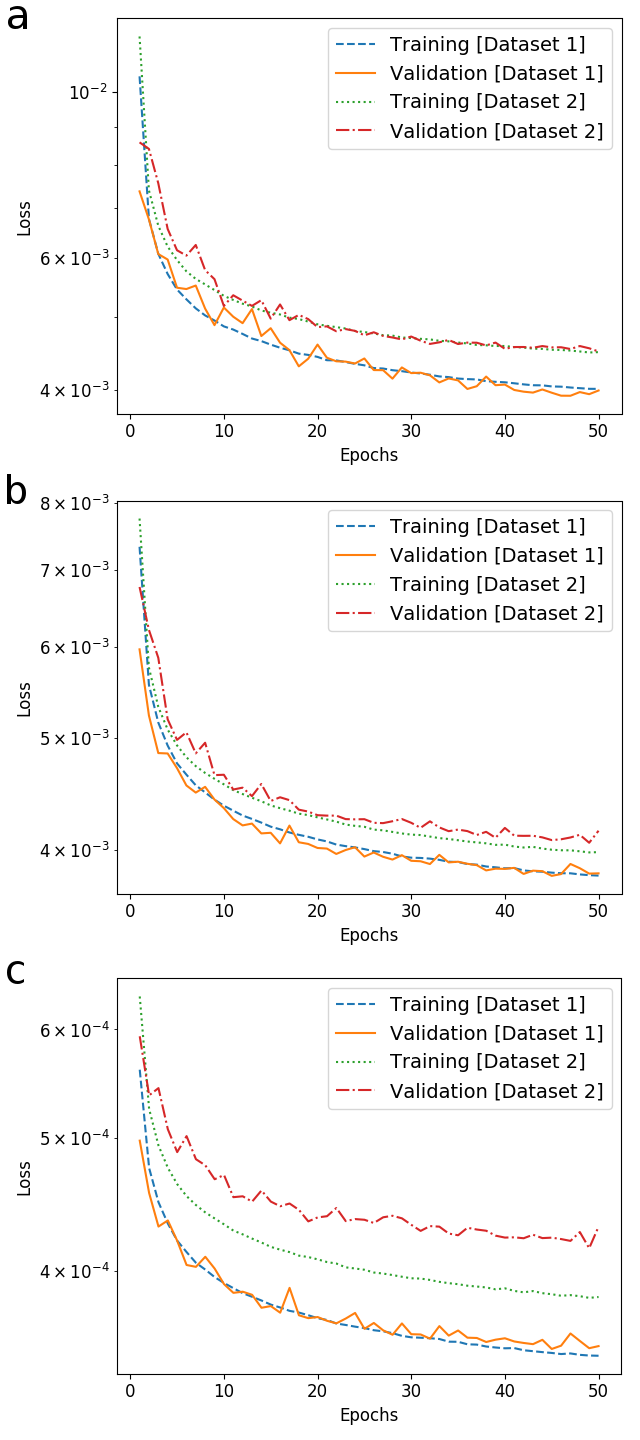}
\caption{ { \bf The mean squared loss for height maps, \vdWSph~and atomic disks.} The mean squared loss on the training and validation sets as functions of
the number of training epochs \textbf{a} for the height map prediction, \textbf{b} for the \vdWSph prediction and \textbf{c} for the atomic disks prediction. }
\label{SI_figLossDSH}
\end{figure*}

The model was trained separately for the three different representations on two different data sets. The first data set contains the elements H, C, N, O, and F (see \reffig{SI_figValidSampl}), and the second data set was extended to additionally contain Si, P, S, Cl, and Br. Table \ref{table:losses} lists the final losses on the trained models and \reffig{SI_figLossDSH} shows the losses as functions of training epochs. The loss has the interpretation of being the square of the average error in height in \AA. The loss on the training and test sets are almost the same in every case except on the second data set with atomic disks. From this we can conclude that the model is not badly overfitting to the training data. The losses on the second data set are greater than on the first one, as would be expected since the data has larger variance.

\begin{table}[b!]
	\centering
	\caption{Losses on the training and test sets for the trained models.}
	\begin{tabular}{l | l l l | l l l }
		 & \multicolumn{3}{c|}{Data Set 1} & \multicolumn{3}{c}{Data Set 2}  \\ \cline{2-7}
		 & Atomic Disks & \vdWSph & Height Map & Atomic Disks & \vdWSph & Height Map \\ \hline
		Train & $3.47 \times 10^{-4}$ &$3.79 \times 10^{-3}$ &$4.01 \times 10^{-3}$& $3.83 \times 10^{-4}$ & $3.98 \times 10^{-3}$ & $4.49 \times 10^{-3}$  \\ 
		Test & $3.50 \times 10^{-4}$& $3.79 \times 10^{-3}$&$3.97 \times 10^{-3}$& $4.41 \times 10^{-4}$ & $4.16 \times 10^{-3}$ & $4.57 \times 10^{-3}$
	\end{tabular}
	\label{table:losses}
\end{table}

\begin{table}[t!]
		\centering
		\caption{{\bf Model architecture}. The factors in parentheses denote parallel layers for separate output branches. The total number of parameters is 122,811.}
		\footnotesize
		\begin{tabular}{r l|l l l l}
			\multicolumn{2}{c|}{Layer type} & Output dimension & Kernel size & Activation & Parameters \\ \hline
			0 & Input & $128 \times 128 \times 10 \times 1$ & - & - & - \\ \hline
			1 & 3D conv & $128 \times 128 \times 10 \times 4$ & $3 \times 3 \times 3$ & $\mathrm{LeakyReLU}_{0.1}$ & 112 \\ \hline
			2 & Avg pool & $64 \times 64 \times 5 \times 4$ & $2 \times 2 \times 2$ & - & -  \\ \hline
			3 & 3D conv & $64 \times 64 \times 2 \times 8$ & $3 \times 3 \times 3$ & $\mathrm{LeakyReLU}_{0.1}$ & 872 \\ \hline
			4 & Avg pool & $32 \times 32 \times 2 \times 8$ & $2 \times 2 \times 2$ & - & -  \\ \hline
			5 & 3D conv & $32 \times 32 \times 2 \times 16$ & $3 \times 3 \times 3$ & $\mathrm{LeakyReLU}_{0.1}$ & 3472 \\ \hline
			6 & Avg pool & $16 \times 16 \times 2 \times 16$ & $2 \times 2 \times 1$ & - & -  \\ \hline
			7 & Reshape to 2D & $16 \times 16 \times 32$ & - & - & - \\ \hline
			8 & 2D conv & $16 \times 16 \times 64$ & $3 \times 3$ & $\mathrm{LeakyReLU}_{0.1}$ & 18496 \\ \hline
			9 & 2D conv & $16 \times 16 \times 64$ & $3 \times 3$ & $\mathrm{LeakyReLU}_{0.1}$ & 36928 \\ \hline
			10 & NN-upsample & $32 \times 32 \times 64(\times 3)$ & - & - & - \\ \hline
			11 & 2D conv & $32 \times 32 \times 16(\times 3)$ & $3 \times 3$ & $\mathrm{LeakyReLU}_{0.1}$ & 9232$(\times 3)$ \\ \hline
			12 & 2D conv & $32 \times 32 \times 16(\times 3)$ & $3 \times 3$ & $\mathrm{LeakyReLU}_{0.1}$ & 2320$(\times 3)$ \\ \hline
			13 & NN-upsample & $64 \times 64 \times 16(\times 3)$ & - & - & - \\ \hline
			14 & 2D conv & $64 \times 64 \times 16(\times 3)$ & $3 \times 3$ & $\mathrm{LeakyReLU}_{0.1}$ & 2320$(\times 3)$ \\ \hline
			15 & 2D conv & $64 \times 64 \times 16(\times 3)$ & $3 \times 3$ & $\mathrm{LeakyReLU}_{0.1}$ & 2320$(\times 3)$ \\ \hline
			16 & NN-upsample & $128 \times 128 \times 16(\times 3)$ & - & - & - \\ \hline
			17 & 2D conv & $128 \times 128 \times 16(\times 3)$ & $3 \times 3$ & $\mathrm{LeakyReLU}_{0.1}$ & 2320$(\times 3)$ \\ \hline
			18 & 2D conv & $128 \times 128 \times 16(\times 3)$ & $3 \times 3$ & $\mathrm{LeakyReLU}_{0.1}$ & 2320$(\times 3)$ \\ \hline
			19 & 2D conv & $128 \times 128 \times 1(\times 3)$ & $3 \times 3$ & None/$\mathrm{ReLU}$/$\mathrm{ReLU}$ & 145$(\times 3)$ \\ \hline
		\end{tabular}
	    \label{table:model2}
	\end{table}

\FloatBarrier
\section{Probe Particle simulations}

For each molecule, the Lennard-Jones and Electrostatic force field was pre-calculated on a regular cubic real-space grid of size $30\times30\times30$ \AA\ with spacing 0.1 \AA\ in each direction and stored in a floating point 3D texture on a GPU. The Lennard-Jones field is calculated using standard OPLS parameters listed in table \ref{Tab:LJparams} (there is no re-fitting of this force field within our approach). The electrostatic field is calculated from Mulliken charges taken from the quantum chemistry simulations (note that the use of high-level couple cluster calculations is not important for getting reasonable charges, but offers high accuracy charge densities for benchmarking and future applications). 

The weighting of molecular orientations in the database mentioned in the main text is done automatically by sorting orientations with respect to distance-weighted counting of atoms using the function $S=\sum_i \exp( - \beta (z_i - z_{close} ) )$. $z_i$ is the z-coordinate of atom $i$ in the coordinate system of the current scan, $z_{close}$ corresponds to the closest atom and the decay factor $\beta$ is currently set to 1.0 [\AA$^{-1}$]. For a constant number of atoms, the function is clearly maximal when all $z_i$ are similar. 

In the process of the neural network training, we vary some simulation parameters in order to regularize the training and to make it less dependent on a particular setup, therefore more robust with respect to uncertain experimental conditions. Predominantly we vary the equilibrium tilt of the probe particle in the range $\pm $1.0\AA, reflecting the asymmetric absorption of CO typical in experiment. We also vary the charge as well as Lennard-Jones radius and binding energy by 0.1e, 0.2 \AA\ and 5$\%$ respectively.

\begin{table}[ht]
\caption{Lennard-Jones parameters in Probe Particle Simulation and rigid body relaxation of surface}
\centering
\begin{tabular}{ c c c }
type & $R_{ii}$[\AA] & $E_{ii}$[eV] \\ 
\hline
H  & 1.4870 & 0.000681 \\ 
C  & 1.9080 & 0.003729 \\
N  & 1.7800 & 0.007372 \\
O  & 1.6612 & 0.009106 \\
S  & 2.0000 & 0.010841 \\
Cu & 2.2300 & 0.010000 
\end{tabular}
\label{Tab:LJparams}
\end{table}

\FloatBarrier

\let\oldthebibliography=\thebibliography
\let\oldendthebibliography=\endthebibliography
\renewenvironment{thebibliography}[1]{%
     \oldthebibliography{#1}%
     \setcounter{enumiv}{67}%
}{\oldendthebibliography}